\documentstyle[aps,epsfig,delarray]{revtex}
\newcommand{\sla}[1]{\mbox{$#1\!\!\!/$}}
\newcommand{\be}{\begin{equation}}
\newcommand{\ee}{\end{equation}}
\newcommand{\bea}{\begin{eqnarray}}
\newcommand{\eea}{\end{eqnarray}}
\newcommand{\hf} {{\frac12}}
\newcommand{\nonu}{\nonumber\\}
\def\eq#1{(\ref{#1})}
\def\la{\langle}
\def\ra{\rangle}
\def\ord#1{{\cal O}(#1)}

\def\mr#1{{\mathrm#1}}
\def\cd#1{{\cal D}[#1]}

\def\ci{{\rm i}}
\def\fd#1#2{{\delta#1\over\delta#2}}
\def\psib{\bar\psi}

\begin{document}

\draft
\title{Periodic ground state for the charged massive Schwinger model}
\author{S. Nagy$^1$, J. Polonyi$^{2,3}$, and K. Sailer$^1$}
\address{$^1$ Department for Theoretical Physics, University of Debrecen,
Debrecen, Hungary}
\address{$^2$ Institute for Theoretical Physics, Louis Pasteur University,
Strasbourg, France}
\address{$^3$ Department of Atomic Physics, Lorand E\"otv\"os University,
Budapest, Hungary}
\date{\today}
\maketitle

\begin{abstract}
It is shown that the charged massive Schwinger model 
supports a periodic
vacuum structure for arbitrary charge density,
similar to the common crystalline layout known in solid state
physics. The dynamical origin of the inhomogeneity is
identified in the framework of the bozonized model and
in terms of the original fermionic variables.
\end{abstract}


\section{Introduction}
An important feature of Quantum Field Theories is the formal separation
of the "active" and "spectator" degrees of freedom. In fact, the conventional
contributions to the perturbation series of a Green function,
represented by Feynman graphs, involve more and more actively
participating particles as the order of the expansion is increased
and the ground state with its infinitely many real or virtual
particles remains formally hidden. This scheme is really efficient
only if the vacuum is "empty" which is usually realized by the
adiabatic turning on and off the interactions as the time
evolves. But serious complications arise in this scheme when
the vacuum is "non-empty", i.e. contains a significant number of
particles. When the constituents of the vacuum form a rigid
system then the vacuum is called solid and space-symmetries
are violated. When the constituents are not localized
then the vacuum can be considered as liquid.

When the fluctuations are sufficiently small then the saddle point
expansion can be used to turn this qualitative picture into
a systematical description. The "non-empty" vacuum consists
of a condensate in this scheme. What is the momentum of the condensed
particles? In case of vanishing momentum the vacuum is homogeneous
and the dynamics of the excitations can be described
in a straightforward manner. But it may happen that
the vacuum is made up by particles of non-vanishing momentum
in which case the saddle point is inhomogeneous and
breaks the space-time symmetries. Depending on the `inertia'
of the saddle point the zero modes arising from this formal
symmetry breaking are either integrated over (liquid) or kept at a
fixed value (solid).

A few examples for liquids are the following. The short range
correlations of the vacuum in Yang-Mills models correspond
to a liquid of localized tree-level saddle points, instantons \cite{inst}.
The one-loop effective action indicates the presence of other
condensates \cite{savv} which must be inhomogeneous in order
to preserve the external and internal symmetries. The mixed phase at
first order phase transitions and the corresponding
Maxwell-cut results from the soft modes which are generated by the
inhomogeneous saddle points of the Kadanoff-Wilson blocking procedure
in renormalizing the action \cite{instab}. The metallic lattice
is the best known example of solids, and similar, periodically
modulated ground state belongs to the Wigner lattice of dilute electron
gas \cite{Wigner 1934} or the charge density wave phase in solids
\cite{gruner}.

Our current understanding of such vacua is severely limited due to
the strong interactions or correlations between the "active" and the
"spectator" particles. This is easy to understand in the framework
of the saddle point expansion. The soft zero modes of the
inhomogeneous saddle points in a liquid are easy to excite and they
usually lead to serious IR divergences in the semiclassical expansion.
There may not be soft modes present in the ground state of a solid
but momenta borrowed from the inhomogeneous condensate generate
nonperturbative phenomena such as the opening of gaps. In addition to
the strong coupling between the "active" and "spectator" degrees of
freedom the dynamical origin of the condensation and the explicit
construction of the ground state from first principles represent a
so far unmatched challenge in both cases. We believe that the
treatment of the soft modes is a more serious and difficult problem
than those of the momentum nonconserving processes. The goal of
the present work is to trace the origin of the periodically modulated
vacuum in one of the simplest interactive theory, Quantum
Electrodynamics in 1+1 dimensions, the Schwinger model
\cite{Schwinger 1951}.

Which part of the (effective) action is responsible of the
inhomogeneity of the vacuum? The inhomogeneity suggests the presence of
strongly distance-dependent interactions in the system and it is
natural to expect that such interactions arise from the
higher derivative terms in the action rather than from the
ultralocal potential energy. Such a relation between
the presence of higher order derivative terms with sufficiently
strong coefficients in the action and the inhomogeneity of the
vacuum has already been confirmed in a number of cases \cite{afvac}.
What was left open by these works is the dynamical origin of the
higher order derivative terms in the effective action
which drive the condensation at nonvanishing momentum.
The higher order derivative terms
which are supposed to be responsible for an eventual inhomogeneity
in QED are to be sought in the effective theory for the photons
or for the density. The simplifications
which occur when we constrain ourselves into 1+1 dimensions
allow us to use simple but powerful analytical and numerical
methods, such as bosonization and the variational approach,
to explore these effective theories in a nonperturbative manner.
The main result of this work is that the vacuum of the
massive Schwinger model in the presence of nonvanishing total
electric charge is periodically modulated. It is reassuring that both the
bosonization and the variational approach yield the same
conclusion. The bosonization allows to identify a mechanism,
more involved as anticipated in the works \cite{afvac}, as the
driving force in forming the periodically modulated vacuum.
The inhomogeneity arises from the competition of an
unusual piece of the kinetic energy which contains the
first power of the space derivative, the boundary conditions
and the periodic part of the potential energy. The periodicity
of the vacuum field configuration in space originates from the
periodicity of a term in the potential energy in the field
variable.

The Schwinger model has already been extensively investigated.
What will be important from the point of view of the present
work is that the confinement of the electric charge has been
established \cite{Schw 1962,Coleman 1975 2,Rothe 1979,Fischler 1979}
and simple analytic considerations hint that the vacuum with nonvanishing
background charge is inhomogeneous \cite{Fischler 1979}.

It is worth mentioning that a non-trivial vacuum structure appears
in QCD$_4$ as well at high fermion densities within the large $N_c$
expansion \cite{Deryagin}. This result motivated the search for
periodic structure in two-dimensional field theoretical models where
the investigation can be carried out without approximations.
It was found that the Gross-Neveu and the 't Hooft models exhibit periodic
baryon density \cite{Thies}, and the multiflavor Schwinger model
and QCD$_2$ also shows up periodic ground state \cite{Christiansen 1997}.

The resemblance of the Schwinger model to a one-dimensional electron system
may one also lead to the idea of the existence of a
periodic ground state. Some compounds can have such
atomic structure that they exhibit one-dimensional metallic properties
and show periodic structure \cite{Comes}. An  analogue of the
Wigner crystal appears in a one-dimensional spin system
with short range, nearest neighbour interaction
\cite{Hubbard 1978}, and  a one-dimensional electron gas with a long
range $U(r)\sim 1/r$ type Coulomb potential also exhibits periodic ground
state \cite{Schulz}. Another indication of the inhomogeneous vacuum
structure in one spatial dimension comes from the non-relativistic
Peierls mechanism \cite{peierls}.

The massless Schwinger model is exactly solvable \cite{Baaquie 1982} and
the explicit computation of the fermion determinant leads to an
effective theory with massive photons and confinement \cite{Coleman 1975 2}.
When the vacuum polarization effects are neglected then the electric
flux conservation induces a flux tube between an electron-positron pair
in the absence of other charges. The resulting linear potential, the
hard confinement mechanism, renders the positronium confined.
Let us now allow the vacuum polarization
to be present and try to separate a member of the positronium, that of the
meson of the Schwinger model. The electric flux tube breaks
up due to electron-positron pair creation when the energy of the
stretched flux tube is sufficiently large and the members of the newly
created pair bind to those of the original pair. This is the soft
confinement mechanism and one ends up with more neutral
mesons again in this manner. The linear potential between the
electron-positron pair becomes saturated by virtual pair creations.
The potential between a pair of static test
charges can easily be obtained in the presence of vacuum polarization
\cite{Coleman 1975 2} and it shows that the total screening, the
soft confinement mechanism, occurs for arbitrary value of the charge.

The massive Schwinger model is not exactly solvable and the
potential between a pair of test charges is periodic function
of the charge with period length given by the elementary charge $e$
and is saturated for integer multiples of $e$ only
\cite{Coleman 1975 2}. The mass gap prevents the vacuum to
screen out non-integer multiples of the elementary charge.
Notice that the massless model is singular in the sense that
arbitrarily small mass is enough to prevent the screening
of non-integer charges at sufficiently large distances. At the end any
charge is confined in the massive model as well but the
integer or non-integer charges are confined by the soft or hard
mechanisms, respectively.

Excitations above a fermionic vacuum with well-defined particle number
are always consisted of particle-hole pairs and are therefore of
bosonic nature. It is the special feature of the 1+1 dimensional world
that the effective theory for these bosonic excitations is local.
The local effective theory resulting from the bosonization of the
massive Thirring model is the sine-Gordon model \cite{Coleman 1975 1}.
These bosonization rules are widely used for the investigation of the
Schwinger model
\cite{Rothe 1979,Fischler 1979,Lowenstein,Coleman 1976,Casher 1974}.
The massless case yields a free scalar theory and the massive
theory leads to the massive sine-Gordon model \cite{Kij}.
The massive Schwinger model was also investigated by bosonization
technique \cite{Kao} and by functional methods \cite{Christiansen}
at non-vanishing chemical potential $\mu$ and temperature $T$.
The existence of a periodic chiral condensate with the wavenumber of
$2\mu$ has been established for arbitrary temperature, too.

Arguments were given in the framework of the tree-level solution
of the bosonized model \cite{Fischler 1979} that the massive Schwinger
model exhibits a periodic ground state in the presence of a static,
homogeneous background charge density.
A more systematic investigation of the inhomogeneity of
the vacuum of the massive Schwinger model in the presence of
homogeneous external charge density, $\rho_{\mr{ext}}$, is presented
in this paper. We attack the problem from two different directions.
First by minimizing the tree-level
expression of the energy functional in the bosonized form of the
model and after that by minimizing the energy with respect to
the parameters of a static, periodic background electric field
in the fermionic form of the theory and by retaining the quantum
fluctuations of the photon field up to the two-loop order.
The results obtained by both approaches are in qualitative agreement.
Namely, the ground state exhibits periodically modulated charge density
with decreasing amplitude for increasing $\rho_{\mr{ext}}$.
For large $\rho_{\mr{ext}}$ numerical calculations failed to be
conclusive regarding the true energy minimum. Analytic considerations
were used in this density regime with the result that
the vacuum remains periodic for arbitrarily large values of
$\rho_{\mr{ext}}$.

The paper is organized as follows. Section II contains the study
of the tree-level bosonized theory in the presence of the
homogeneous external charge density $\rho_\mr{ext}$.
The minimum of the energy functional of the model is found
by numerical minimization of the classical vacuum energy.
For large values of $\rho_\mr{ext}$ when the result is more unstable
with respect to numerical errors the stability of the periodic vacuum
has been shown analytically, by expanding the tree-level energy
in powers of the amplitude of charge density wave in the vacuum.

The fermionic investigations are presented in Section III.
Since integer charges are screened and do not leave behind long range
photon field \cite{Fischler 1979} the perturbation expansion in $e$
is reliable by using the original fermionic and photonic degrees of
freedom. We follow a variational strategy and minimize the energy
of the vacuum as the function of the induced photon field in the vacuum.
The fermionic degrees of freedom are integrated out in the presence
of a static, sinusoidal electric field and the quantum fluctuations of
the photon field are taken into account up to two-loop
diagrams for the energy. The external charge density $\rho_\mr{ext}$
is introduced indirectly via a fermionic chemical potential $\mu$.
The energy of the vacuum is finally minimized with respect to the amplitude
and the wavelength of the static periodic background electric field.
The numerical minimization procedure finds the periodic ground state
energetically favorable as compared to the homogeneous one below certain
value of $\rho_\mr{ext}$. The problem of the high density regime is of the
same origin as in the bosonized study, namely the smallness of the
amplitude of the induced photon field in the vacuum.
An analytic calculation in the framework of the perturbation expansion
in the amplitude of the induced photon field predicts a periodic
ground state even in this density region.

Finally, the conclusion is drawn up in Sect. \ref{summary}.
Appendices \ref{DH} and \ref{enspec}
present briefly the numerical solutions of the
Dirac equation in periodic background potential and the band
structure of the fermionic spectrum, respectively. Explicit
expressions for the Feynman diagrams for the energy and charge
densities up to the two-loop order
are given in Appendix \ref{app:en}, and the details of the numerical
search for the energy minimum are given.

\section{Minimization of the energy in the bosonized model}
This section contains the tree-level determination of the vacuum
structure of the bosonized model.

\subsection{Hamiltonian}
The Lagrangean of the  massive Schwinger model is given as
\be
{\cal L}=-\frac14 F_{\mu\nu}F^{\mu\nu}+\bar\psi\gamma^\mu(\partial_\mu-
\ci e A_\mu)\psi-m\bar\psi\psi,
\label{mSch}
\ee
where $F_{\mu\nu}=\partial_\mu A_\nu-\partial_\nu A_\mu$, $m$ and $e$ are the
bare rest mass of the electron and  the bare coupling constant, respectively.
The bosonization rules are \cite{Fischler 1979}:
\bea
:\bar\psi\psi:&\to~-cmM\cos(2\sqrt{\pi}\phi),~~~~~~
:\bar\psi\gamma_5\psi:&\to~-cmM\sin(2\sqrt{\pi}\phi),\nonu
j_\mu=:\bar\psi\gamma_\mu\psi:&\to~\frac1{\sqrt{\pi}}
\varepsilon_{\mu\nu}\partial^\nu\phi,~~~~~~~~~~~~~~~~~
:\bar\psi\ci\sla\partial\psi:&\to~\frac12 N_m (\partial_\mu\phi)^2,
\eea
where $N_m$ denotes normal ordering with respect to the fermion mass $m$,
$c=\exp{(\gamma)}/2\pi$ with the Euler constant $\gamma$,
and $M=e/\sqrt{\pi}$ the `meson' mass. It is believed
that the presence of a non-vanishing background charge density does not
affect these transformation rules \cite{Kao}.
The Hamiltonian of the system in Coulomb gauge is given by
\be
{\cal H}=\int_x \bar\psi_x(\ci \gamma_1\partial_1+m)\psi_x
-\frac{e^2}{4}\int_{x,y} j_{0,x}|x-y|j_{0,y},
\label{hamcoul}
\ee
with $\int_x=\int_0^T dx^0 \int_{-L}^L dx^1$.
According to the bosonization rules this Hamiltonian is equivalent to those
of the massive sine-Gordon model,
\be
{\cal H}[\Pi,\phi]=N_m\int_x \left[\frac12\Pi^2_x+\frac12(\partial_1\phi_x)^2+
\frac12M^2\phi^2_x-cmM\cos(2\sqrt\pi\phi_x)\right]
\label{hamsc}
\ee
where $\Pi_x$ denotes the momentum variable canonically conjugated
to $\phi_x$.

Our purpose is to determine the vacuum of the massive Schwinger model
in the presence of an external static particle density $\rho_{\mr{ext}~x}$
which is added to the density $j_{0,x}$ in Eq. (\ref{hamcoul}),
\be
{\cal H}_\mr{ext}=\int_x \bar\psi_x(\ci \gamma_1\partial_1+m)\psi_x
-\frac{e^2}{4}\int_{x,y} (j_{0,x}+\rho_{\mr{ext}~x})|x-y|(j_{0,y}+
\rho_{\mr{ext}~y}).
\label{hamcoulch}
\ee
The external charge is represented by the external field
$\phi_{\mr{ext}~x}$ in the bosonized Hamiltonian (\ref{hamsc}) as
\be
{\cal H}_\mr{ext}[\Pi,\phi]=
N_m\int_x\left[\frac12\Pi^2_x+\frac12(\partial_1\phi_x)^2+
\frac12M^2(\phi_x+\phi_{\mr{ext}~x})^2-cmM\cos(2\sqrt\pi\phi_x)\right]
\label{hamscch}
\ee
where
\be
\rho_{\mr{ext}~x}=\frac1{\sqrt{\pi}}\partial_1\phi_{\mr{ext}~x}.
\ee
The external particle density is assumed to be static and constant
in the interval $x^1=z\in[-L,L]$ and vanishing elsewhere,
therefore we write $\phi_{\mr{ext}~x}=b z$ when $|z|\le L$ and
$\phi_{\mr{ext}~x}=0$ elsewhere for any $x^0=t$.
It is advantegous to introduce the field variable
\be
\tilde\phi_z=\phi_z+bz
\ee
which allows us to write the total particle density as
\be
\rho_x = \frac1{\sqrt\pi}\partial_1\tilde\phi_x.
\label{cd}
\ee

The tree-level vacuum can be constructed by minimizing the
Hamiltonian \eq{hamsc} as the functional of the static field
configuration $\phi_x$ with $\Pi_x=0$. The minimum is reached at
$\phi_{\mr{gr}~x}=\la0|\phi_x|0\ra$ and the value of the Hamiltonian
at this field configuration, $E(b)={\cal H}_\mr{ext}[0,\phi_\mr{gr}]$,
can be identified by the tree-level vacuum energy. Lattice regularization
of the Hamiltonian for static field,
\be
{\cal H}_\mr{ext}[0,\phi]=\int_z\left[\frac12(\partial_1\phi_z)^2
+\frac{e^2}{2\pi}(\phi_z+b z)^2
-\frac{cme}{\sqrt{\pi}}\cos(2\sqrt{\pi}\phi_z)\right],
\label{Ener}
\ee
yields
\be
a{\cal H}_L[0,\phi]=\hf\sum_{n=0}^N(\phi_{n+1}-\phi_n)^2
+\frac{e_L^2}{2\pi}\sum_{n=0}^N(\phi_n+bz_n)^2
-\frac{cm_Le_L}{\sqrt{\pi}}\sum_{n=0}^N\cos(2\sqrt{\pi}\phi_n),
\label{Hamdisc}
\ee
where $a$ stands for the lattice spacing, $e_L=ea$, $m_L=ma$,
$z_n=z_0+na$, $a=2L/(N+1)$, $z_0=-L$, $z_{N+1}=L$ and
$\phi_n=\phi_{z_n}$. The boundary conditions
\be
\phi_0=\phi_{N+1}=0
\label{constraint}
\ee
have been used in order to restrict the computation into the sector
with vanishing induced charge.

In order to understand the origin of the periodic structure of the
vacuum we rewrite the static Hamiltonian \eq{Hamdisc}
in terms of the shifted variable $\tilde\phi$ as
\be\label{frko}
a\tilde{\cal H}_L[0,\tilde\phi]=\hf\sum_{n=0}^N
(\tilde\phi_{n+1}-\tilde\phi_n-\sqrt{\pi}\rho_{\mr{ext}~L})^2
+\frac{e_L^2}{2\pi}\sum_{n=0}^N\tilde\phi_n^2
-\frac{cm_Le_L}{\sqrt{\pi}}\sum_{n=0}^N
\cos(2\sqrt{\pi}(\tilde\phi_n-bz)),
\ee
where $\rho_{\mr{ext}~L}=a\rho_\mr{ext}=ab/\sqrt{\pi}$ denotes the amount
of external particles distributed between two consecutive lattice sites.
This expression reveals a competition in forming the vacuum,
taking place between the kinetic and the potential energies,
the first and the remaining terms on the right hand side of
Eq. \eq{frko}. Let us first ignore for simplicity the quadratic
mass term and the shift $-bz$ in the argument of the cosine
function on the right hand side, a simplification which yields
the Hamiltonian of the Frenkel-Kontorova model \cite{frko}. The tree-level
vacuum of this model produces infinitely many commensurate-incommensurate
transitions and displays a rather involved phase structure with the
devil-staircase feature due to the competition between two
dimensionless parameters \cite{frkoph}. In fact, the kinetic energy prefers
\be\label{kinepref}
\tilde\phi_n=\mr{const}.+n\sqrt{\pi}\rho_{\mr{ext}~L}
\ee
and the potential energy is minimal for $\tilde\phi_n=j\sqrt{\pi}$
(with $n$ and $j$ integers) and the
vacuum is trivial, i.e. a linear function of the coordinate,
for integer $\rho_{\mr{ext}~L}$ only. The vacuum of the complete Hamiltonian
\eq{frko} is always the result of a compromise between the kinetic
energy with the preference expressed by Eq. \eq{kinepref}
and the potential energy which prefers $\tilde\phi=0$ and
$\tilde\phi_n=\sqrt{\pi}(j-n\rho_{\mr{ext}~L})$. The competition
between the kinetic and the potential energy is never trivial
due to the quadratic mass term
but is at least simplified when $\rho_{\mr{ext}~L}$ is integer.
For sufficiently small lattice spacing $\rho_{\mr{ext}~L}<1$
each energy expression on the right hand side of Eq. \eq{frko}
enters in the competition for the vacuum.

Such an involved vacuum structure
is characteristic of the tree-level solution in lattice regularization
only. The quantum fluctuations should smear most of the
commensurate-incommensurate transitions out. A similarly smeared behaviour
is what one finds when the cutoff is ignored in the tree-level sector,
i.e. when the minimum energy configuration is searched in the naive,
classical continuum limit of Eq. \eq{frko},
\be\label{sener}
\tilde{\cal H}_\mr{ext}[0,\tilde\phi]=\int_z\left[
\frac12(\partial_1\tilde\phi_z-\sqrt{\pi}\rho_\mr{ext})^2
+\frac{e^2}{2\pi}\tilde\phi_z^2
-\frac{cme}{\sqrt{\pi}}\cos(2\sqrt{\pi}
(\tilde\phi_z-\sqrt{\pi}\rho_\mr{ext}z))\right],
\ee
subject of the boundary conditions $\tilde\phi_{-L}=0$,
$\tilde\phi_L=2\sqrt{\pi}L\rho_\mr{ext}$. In the Frenkel-Kontorova
limit when the mass term and the $z$-dependence in the argument
of the cosine function are ignored then $\phi_z$ develops
oscillatory structure in the vacuum which changes smoothly
with the parameters of the model. In fact, the kinetic energy
prefers to distribute the total change
$\tilde\phi_L-\tilde\phi_{-L}=2\sqrt{\pi}L\rho_\mr{ext}$
in a linear manner but the potential energy introduces a periodic
modulation. The period length can be determined by noting that
$\tilde\phi$ should change by $\sqrt{\pi}$ within a period.
Such a simple argument gives the period length
$\ell_0=1/\rho_{ext}$ for small $me$. The fermions correspond to kinks
of the sine-Gordon model according to the bosonization therefore it
is not surprising to find that there is just one particle per period in
such a vacuum state. The $z$-dependent shift in the cosine function
takes out the driving linear term from $\tilde\phi$ but tends
to generate periodic oscillations with the same period length.

Notice that the source of the inhomogeneity of the vacuum is an
unusual, $\ord{\partial_1}$, gradient term in the kinetic energy.
This contribution to the energy, together with the
boundary conditions and the periodic potential energy
form the periodic modulation in the vacuum. Furthermore the
space inversion symmetry is broken explicitly by the
$\ord{\partial_1}$ term and the boundary conditions. It is interesting
to compare this situation with those encountered in earlier
studies \cite{afvac} where the dispersion relation of the form
\be
\epsilon(p)=C_6p^6+C_4p^4+\frac{p^2}{2}+C_0
\ee
was used with $C_6,-C_4>0$ with non-periodic potential energy
and the tree-level vacuum is expected to be periodic if there
is a region in the momentum space where $\epsilon(p)<0$.
The space-time inhomogeneities are therefore generated by the
competition of terms with different orders of the gradient
only and the space inversion symmetry is broken spontaneously.

\subsection{Numerical results}
The energy minimum was searched by the conjugate
gradient method which started from a number of initial conditions
for $\phi_{z_i}$ and the field configurations corresponding to
the lowest energy only have been singled retained.
The charge densities were then calculated according to Eq. \eq{cd}.
We used $L=16\pi$, $N=800$, $m=0.5;2;5$ and $b\in [0.3;7]$ in the
numerical studies.

The minimum of the expression (\ref{Ener}) was found at $\phi_z=0$ for
vanishing external charge density,
i.e. for $b=0$. The increase of $b$ gave two distinct regions,
separated by a size-dependent point $b=b_L$.

\underbar{$b<b_L$}: For $b$  close to zero one expects that
$\phi_z$ is small and the sinusoidal potential in the equation of motion,
\be
\partial_1\phi_z=\frac{e^2}{\pi}(\phi_z+b z)+2cme\sin(2\sqrt{\pi}\phi_z),
\label{eqmot}
\ee
can well be approximated by the first term of its Taylor series
\be
\partial_1\phi_z\approx\frac{e^2}{\pi}(\phi_z+b z)+4cme\sqrt{\pi}\phi_z.
\label{eqmotlin}
\ee
Such a linearized equation of motion together with the boundary
conditions (\ref{constraint}) yields
\be
\phi_z=b_s L\frac{\sinh(\kappa z)}{\sinh(\kappa L)}-b_s z,~~~
\kappa=\sqrt{\frac{e^2}{\pi}+4cme\sqrt{\pi}},~~~
b_s=b \frac{e^2}{\kappa^2\pi}.
\label{emsollin}
\ee
This solution, shown in Fig. \ref{a0.02}, contains three spatial
regions in the interval $[-L,L]$ and in the longest, central
region $\phi_z$ is linearly decreasing function with the slope $-b_s$.
The analytic results for the slope are in very good agreement with
those obtained numerically. The linear decrease of $\phi_z$ in the
central region describes the partial screening of the external charge
density. In the two other regions, close to the boundaries
at $-L$ and $L$, $|\phi_z|$ approaches abruptly zero.
For every choice of $L$ a critical $b_L$ value was found where the
linear approximation fails to work and the higher-order terms of
the sine function are needed in the equation of motion (\ref{eqmot}).
It was found that the slope $b_s$ reaches $b$ at this point.
The $L$-dependence is $b_L\approx L^{-1.41}$ according to Fig. \ref{linc},
therefore $b_L\to0$ and this type of solutions disappears in the
thermodynamic limit.

\underbar{$b_L<b$}: The increase of $b_s$ to $b$ indicates the
complete screening of the external charge density in the central region.
Furthermore the numerical solution, depicted in Fig. \ref{a0.5m1g1},
reveals an additional periodic structure in $\phi_z$,
$\tilde\phi_z$ is a periodic function of wavelength $\ell$,
$\tilde\phi_z=\tilde\phi_{z+\ell}$. The wavelength $\ell$ and the
amplitude $A$ of $\tilde\phi_z$ were defined numerically
as the distance of the neighbouring zeros of $\tilde\phi_z$ and
the arithmetic average of the magnitude $|\tilde\phi_z|$ at the extrema
of the periodic component, respectively.
Both $\ell$ and $A$ decrease with increasing $b$ in this region as
shown in Figs. \ref{g1loga} and \ref{g1logl}.
This feature opens the possibility of applying
the perturbation expansion in the amplitude $A$ of the induced
periodic field in the vacuum in the
limit of asymptotically large charge densities, $\rho\to\infty$.
Based on Fig. \ref{a0.5m1g1} the periodic part of the scalar field
is approximated by
\be
\tilde\phi_z=A\sin\left(\frac{2\pi}{\ell} z\right).
\ee
By inserting this expression into Eq. (\ref{Ener}) one finds the
energy density
\bea
{\cal E} (A,b) \equiv \frac{E(b)}{2L}&=&
\frac{A^2\pi^2}{\ell^2}+\frac{\pi}{2\ell^2}+\frac{e^2A^2}{4\pi}
-\frac{cme}{\pi^{1/2}} J_1(2\sqrt{\pi}A),
\label{enerres}
\eea
where $J_1(x)$ is the Bessel function of the first kind. Fig. \ref{g1loga}
shows that the amplitude $A$ decreases with increasing  charge density.
Therefore it is sufficient to consider the expression on the
right hand side of Eq. (\ref{enerres}) only up to the order
${\cal O}(A^2)$ for large $\rho\to\infty$. Due to the relation
\be
J_1(x)\approx \frac{x}2  +{\cal O}(x^3)
\ee
valid for small $x$ the energy density takes the form
\bea
{\cal E}(A,b)=\frac{b^2}2+\frac{A^2}4\left(4\pi b^2+\frac{e^2}{\pi}\right)-cmeA
\eea
having a non-trivial minimum at
\bea
A&=&\frac{2cme}{4\pi b^2+\frac{e^2}{\pi}} >0
\label{aext}
\eea
where the Casimir energy is negative,
\bea
{\cal E}(A,b)-{\cal E}(A=0,b)&=&
-\frac{(cme)^2}{4\pi b^2+\frac{e^2}{\pi}}<0.
\eea
Thus one concludes that the ground state of the massive Schwinger model
is periodic for large external charge densities. It is shown in Fig.
\ref{g1loga} that the analytic result of Eq. \eq{aext} is in good agreement
with the numerical one for the charge-dependence of the amplitude $A$.

The relation $\ell=1/\rho_\mr{ext}$ displayed by Fig. \ref{g1logl}
reflects the fact that charges which are integer multiples of $e$ are
completely screened. In fact, as argued in Ref. \cite{Fischler 1979},
the introduction of the charges $\pm e$ at the boundaries corresponds
to the shift $z\to z+\delta z$ with $|\delta z|=\sqrt{\pi}/b$.
This is a symmetry of the vacuum if $\tilde\phi_z$ has the length
of period $\ell=|\delta z|/\nu$ where $\nu$ is integer. According
to the numerical results $\nu=1$ at the energy minimum.
Similar periodic structure is found in Wigner crystals
of itinerant electrons, in certain spin systems \cite{Hubbard 1978}
and in the charge density wave states. The periodicity
usually gives way to homogeneity when
the external charge density is increased because the
overlap integrals between the neighbouring lattice sites increase.
This is not what happens in the massive Schwinger model, where the
simple, leading order perturbation expansion given above shows
that the ground state
keeps its periodicity for arbitrarily large charge densities.

Our conclusion is that in the tree-level approximation of the bosonized
theory the massive Schwinger model has a single periodic phase in
the thermodynamic limit and the homogeneous external charge density is
neutralized in average by a periodic, induced charge density. Integer charges
are completely screened as argued in \cite{Fischler 1979}.

\section{Variational minimization of the energy for QED$_{1+1}$}
Let us consider now the massive Schwinger model in terms of the
original fermionic degrees of freedom and subject to
periodic boundary conditions at the endpoints of a finite spatial
interval. The finite charge density is now introduced by the
chemical potential $\mu$. The system of electrons is easier to polarize
than the 'empty' vacuum and accordingly there is no gap in the free
electron excitation spectrum for $\mu>m$. The photon polarization tensor
is non-vanishing at the Fermi level therefore the Debye screening renders
the photon propagator short ranged and the Coulomb potential vanishing
for large separation \cite{Nagy 2 2002}. Our computation performed
in this formalism supports the results obtained in the bosonized
theory, namely that even  an arbitrarily weak interaction among the
electrons is sufficient to form a periodic ground state. The dynamical
origin of the modulated ground state is the opening of a gap around
the Fermi level.

We are confronted by two complications in describing the vacuum.
First, the confinement of charge renders the fermionic
excitation spectrum non-physical and ill-defined. As discussed
above in the framework of the bosonized theory integer multiples
of the elementary charge are screened by vacuum polarization
at finite charge densities and their Green function is short ranged.
Since only integer charges can be created in the fermionic theory
we expect no problems with perturbation expansion at finite density.
The second problem, the possibility of dynamical generation of
coherent photons, i.e. a background field in the vacuum is more
difficult and has to be handled in a self-consistent manner.
For this end we introduce an external photon field,
\be\label{ansatz}
\bar A^\nu(x^1)=\delta^\nu_0a\cos\left(Qx^1\right),
\ee
with $a\ge 0$ chosen to be a single plane wave for the sake of simplicity.
Since there is only one non-vanishing component of the field strength tensor
$F_{\mu\nu}=\partial_\mu A_\nu-\partial_\nu A_\mu$, such a background
field represents a generic sinusoidal external field in $1+1$ dimensions.
The energy density will be computed in the order $\ord{e^4}$ in the vacuum
for a given $\mu$ and  minimized with respect to the external field,
the variational parameters $a$ and $Q$.

The numerical minimization of the vacuum energy density with respect
to the background field shows that the system manages to lower
the vacuum energy below the `empty', perturbative value by opening
a gap and generating a photon condensate \eq{ansatz} in the vacuum
for small densities. For large densities the perturbative treatment of the
dependence of the vacuum energy on the field \eq{ansatz} is reliable and
yields similar results.
Our analysis does not cover the intermediate density
regime where the density is large enough to make the numerical minimization
of the two-loop energy expression unreliable but small for the application
of the perturbation expansion in $a$.

\subsection{Background field as collective coordinate}\label{genfunc}
The background field is introduced  by the collective
coordinate method into the generating functional for the Green functions.
The vacuum-to-vacuum amplitude of the model is expressed by the path integral
\be
Z=\int\cd{\psib}\cd{\psi}\cd{A}
e^{\ci S_\mr{EM}[A]+\ci S_\mr{D}[A,\psib,\psi]},
\ee
where the action for the photon field $A^\mu$ in Feynman gauge,
\be
S_\mr{EM}[A]=-\frac{1}{4}\int_x F^{\mu\nu}F_{\mu\nu}-
\hf\int_x(\partial^\mu A_\mu)^2=\hf A\cdot D^{-1}\cdot A,
\ee
is expressed in terms of the inverse of the free photon propagator
\be\label{frphprop}
(D^{-1})^{\mu\nu}_{xy}=g^{\mu\nu}\Box_x\delta_{x,y},
\ee
and the Dirac action,
\be
S_\mr{D}[A,\psi,\psib]=\psib\cdot G^{-1}(A)\cdot\psi,
\ee
is given by means of the inverse fermion propagator
\be\label{fermionprop}
G^{-1}(A)=i\gamma^\mu(\partial_\mu-\ci eA_\mu)-m
=\gamma^0i\partial_0-H_D(A)
\ee
with  the Dirac Hamiltonian
\be\label{dirham}
H_D(A)=\gamma^0(-i\gamma^1\partial_1+m-e\gamma^\mu A_{x\mu}).
\ee
We use the notation $\int_x=\int_0^T dx^0 \int_0^L dx^1$,
$f\cdot g=\int_xf_xg_x$ and shall consider the limit $LT\to \infty$ below.

The vacuum of the model will be constructed by means of a variational
method. We introduce an external background field $\bar A^\nu_x$ and
separate the quantum fluctuations $\alpha^\nu_x$,
$A^\nu_x={\bar A}^\nu_x+\alpha^\nu_x$.
The dependence on the external field is retained by the method of collective
coordinates which implies the insertion of the identity
\be
1=\int d\sigma\delta (C[\bar A,\alpha]+\sigma)
\ee
into the path integral,
\be
Z=\int d\sigma Z_\sigma,~~~~
Z_\sigma=\int\cd{\psib}\cd{\psi}\cd{\alpha}\delta(C[\bar A,\alpha]+\sigma)
e^{\ci S_\mr{EM}[\bar A+\alpha]+\ci S_\mr{D}[\bar A+\alpha,\psib,\psi]}
\ee
with
\be
C[\bar A,\alpha]=\frac{1}{4}(F-\bar F)\cdot\bar F
=\hf\alpha\cdot D^{-1}\cdot\bar A .
\ee
The fluctuations of the collective coordinate $\sigma$ are suppressed
in the thermodynamic limit because the background field is
extended and the $\sigma$-integration can be performed
by expanding $\ln Z_\sigma$ around its maximum. The contribution
of the collective coordinate to the vacuum energy density will be
negligible in the thermodynamic limit and the collective
coordinate can be frozen at the maximum as far as the energy
density in the vacuum is concerned.

One usually employs the effective action formalism in similar problems.
There the external source, coupled linearly to the fluctuating field
is supposed to stabilize the vacuum with the desired condensate.
The minimization of the effective action guarantees that the
external source plays no role in the true vacuum. The complication
which renders this method rather involved beyond the leading order
of the loop expansion is the Legendre transformation.
The procedure outlined above leads to simpler expressions
in the two-loop order. Both methods are useful in the case
of stable ground state only. Large amplitude fluctuations
appear in the mixed phase which make the computation of the
convex effective action and the taking into account the
fluctuations of  the collective coordinate difficult.

It will be useful to introduce the generating functional
\bea\label{genfdf}
Z[j,\bar\zeta,\zeta]&=&\int d\sigma Z_\sigma[j,\bar\zeta,\zeta],\nonu
Z_\sigma[j,\bar\zeta,\zeta]&=&\int d\lambda
\int\cd{\psib}\cd{\psi}\cd{\alpha}
e^{\ci S_\mr{EM}[\bar A+\alpha]+\ci S_\mr{D}[\bar A+\alpha,\psib,\psi]
+\ci\lambda(C[\bar A,\alpha]+\sigma)
+\ci j\cdot\alpha +\ci\bar\zeta\cdot\psi+i\psib \cdot \zeta},
\eea
where the constraint is represented as a Fourier integral over $\lambda$.
The generating functional can be written in the perturbation expansion as
\be
Z_\sigma[j,\zeta,\bar\zeta]=\sum_{n=-\infty}^\infty\frac{1}{n!}
\left(\ci e\int_x\fd{}{\zeta_\alpha^x}\gamma^\mu_{\alpha\beta}\fd{}{j_\mu^x}
\fd{}{\bar\zeta_\beta^x}\right)^n Z_{0~\sigma}[j,\zeta,\bar\zeta],
\ee
where
\bea
Z_{0~\sigma}[j,\zeta,\bar\zeta]&=&\exp\Biggl[
\mr{Tr}\ln G^{-1}(\bar A)-\ci\bar\zeta\cdot G(\bar A)\cdot\zeta
-\frac{4\ci}{a^2 Q^2 LT}\sigma^2-\left(2\ci
+4\ci\frac{\bar A{\cdot}j}{a^2Q^2 LT}\right)\sigma\nonu
&&-\frac12{\rm Tr}\ln D^{-1}
-\frac12\ln\left(-\frac{Q^2}4\bar A\bar A\right)
+\frac{\rm i}2 Q^2\bar A{\cdot}\bar A-\frac{\rm i}2j\cdot D'\cdot j\Biggr]
\eea
with
\be\label{modpro}
D^{\prime\mu\nu}_{xy}=D^{\mu\nu}_{xy}-\frac{\bar A_y^\mu\bar A_x^\nu}
{Q^2\bar A\cdot\bar A}.
\ee
The photon propagator \eq{modpro} tends to the free photon propagator
in the thermodynamic limit when the fluctuations parallel to
the background field vanish and we continue using the original photon
propagator $D$.
Finally, the expectation value of an operator $\hat {\cal O}[\bar A]$ is
determined as
\begin{equation}
\label{expvalO}
{\cal O}[\bar A] =\left.
\frac1{Z_\sigma[j,\zeta,\bar\zeta]}\int d\lambda
\int\cd{\psib}\cd{\psi}\cd{\alpha}\hat{\cal O}[\bar A]
e^{\ci S_\mr{EM}[\bar A+\alpha]+\ci S_\mr{D}[\bar A+\alpha,\psib,\psi]
+\ci\lambda (C[\bar A,\alpha]+\sigma)+\ci j\alpha+\ci\bar\zeta\psi+
\ci\bar\psi\zeta}\right|_{j=\zeta=\bar\zeta=\sigma=0},
\end{equation}
where the field variables in the operator $\hat{\cal O}[\bar A]$ are
replaced by functional derivatives with respect to the
corresponding external sources.

\subsection{Energy and charge of the  vacuum}\label{thedyn}

It may happen that the system prefers energetically a periodic
ground state rather than the normal, homogeneous one. Then it
should adjust itself by building up a static, periodic electric field
and the corresponding band structure with the Fermi-level placed
in a forbidden band. In order to decide whether such a readjustment
of the vacuum takes place one has to compare the energy
densities of the homogeneous and modulated vacua. The fermion spectrum
is needed for the determination of the energy and charge densities.
It has been
calculated along with the corresponding eigenspinors by solving the Dirac
equation numerically (see App. \ref{DH} and also App. \ref{enspec} for a
detailed discussion of the fermion spectrum). The fermion Green functions
were constructed according to their Lehmann expansion. The poles on the
complex energy plane were shifted according to the rules established in
\cite{Anishetty 1984} to take into account the chemical potential $\mu$.
The energy density ${\cal E}[\bar A]$ of the system is given by
the expectation value of the 00th component of the energy-momentum tensor.
At the two-loop order this expectation value can be
represented diagrammatically as
\be
{\cal E}[\bar A]= \frac14 a^2Q^2+
\begin{array}{l}\psfig{file=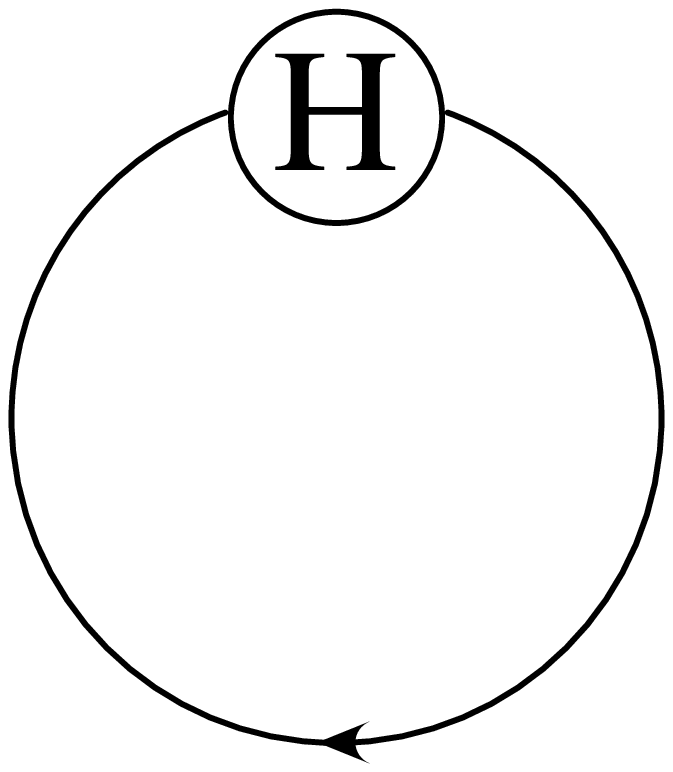,width=1cm}\end{array}
-\ci\begin{array}{l}\psfig{file=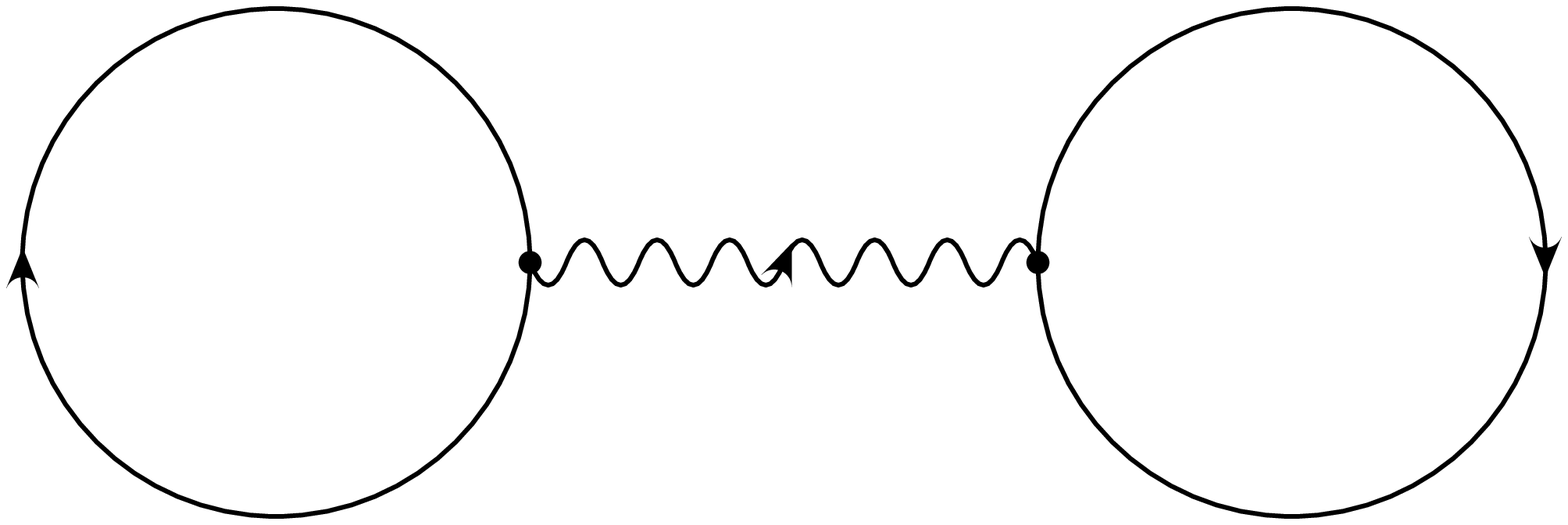,width=1.5cm}\end{array}
+\ci\begin{array}{l}\psfig{file=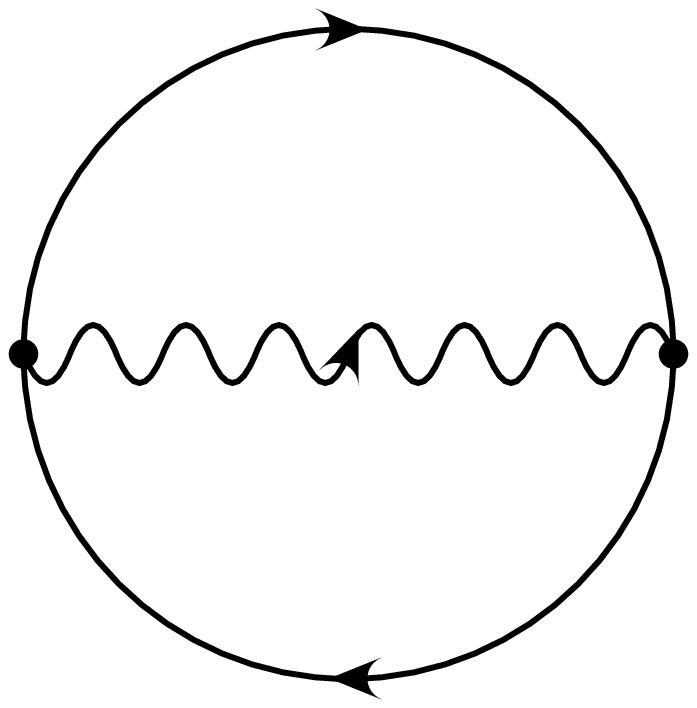,width=1cm}\end{array}
+\begin{array}{l}\psfig{file=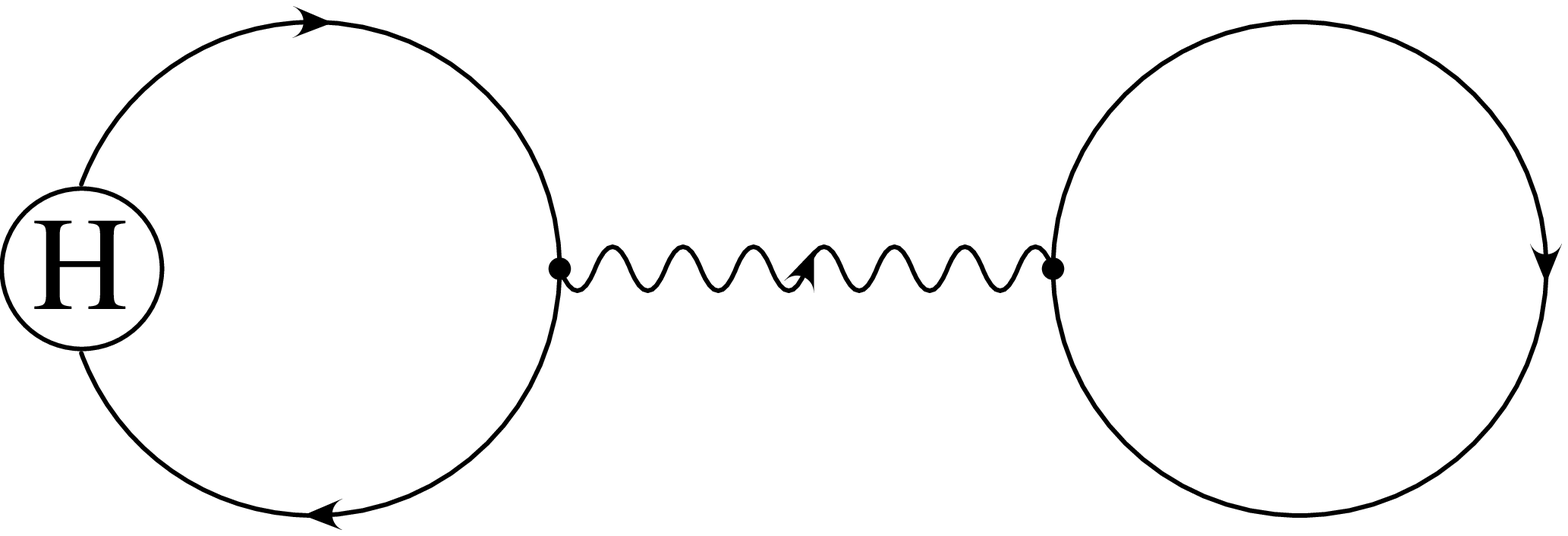,width=1.5cm}\end{array}
-\begin{array}{l}\psfig{file=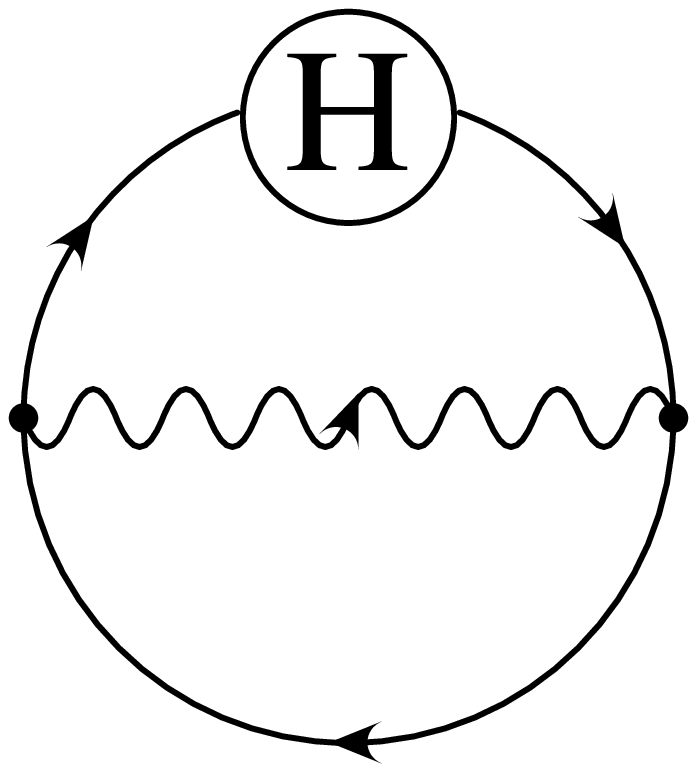,width=1cm}\end{array}
\label{1-2loop},
\ee
the details of the calculation are given in  App. \ref{2lform}.
There is no need of mass and charge renormalizations in QED$_{1+1}$, but
UV and IR divergences appear in the diagrams, which should be properly
handled, see App. \ref{UVIR}.
We introduce the Casimir energy which is usually the energy
difference of the states with and without a classical object and was
investigated thoroughly \cite{Bordag} by the collective coordinate method.
In our case it is the background field which plays the
role of the classical object and the Casimir energy is
\be\label{omcas}
{\cal E}_\mr{C}(a,Q,\mu)={\cal E}_\mr{per}(a,Q,\mu)-{\cal E}_\mr{n}(0,0,\mu).
\ee
In order to understand the structure of the vacuum, we  need another
important observable, the average charge density $\rho$. Its two-loop
order expectation value is
\be
\rho[\bar A]=\begin{array}{l}\psfig{file=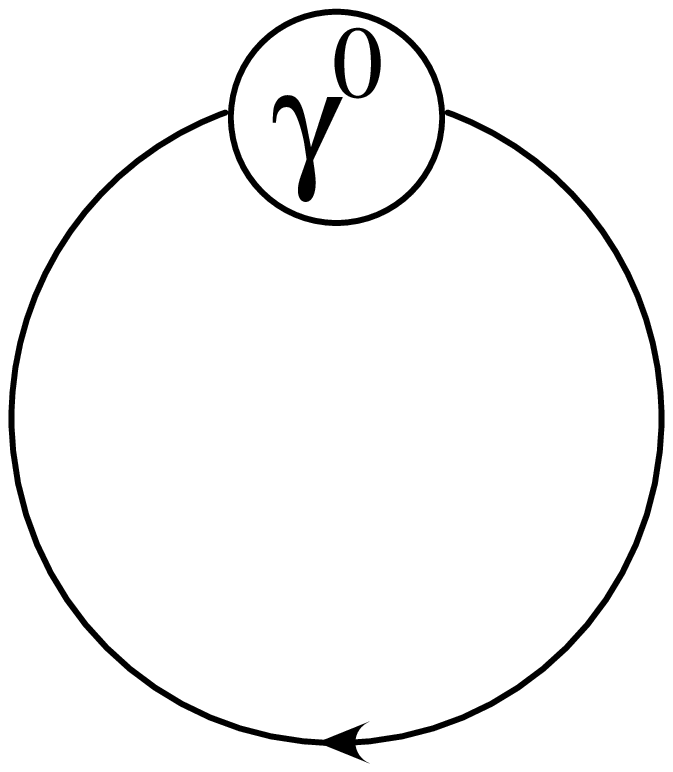,width=1cm}
\end{array}
+\begin{array}{l}\psfig{file=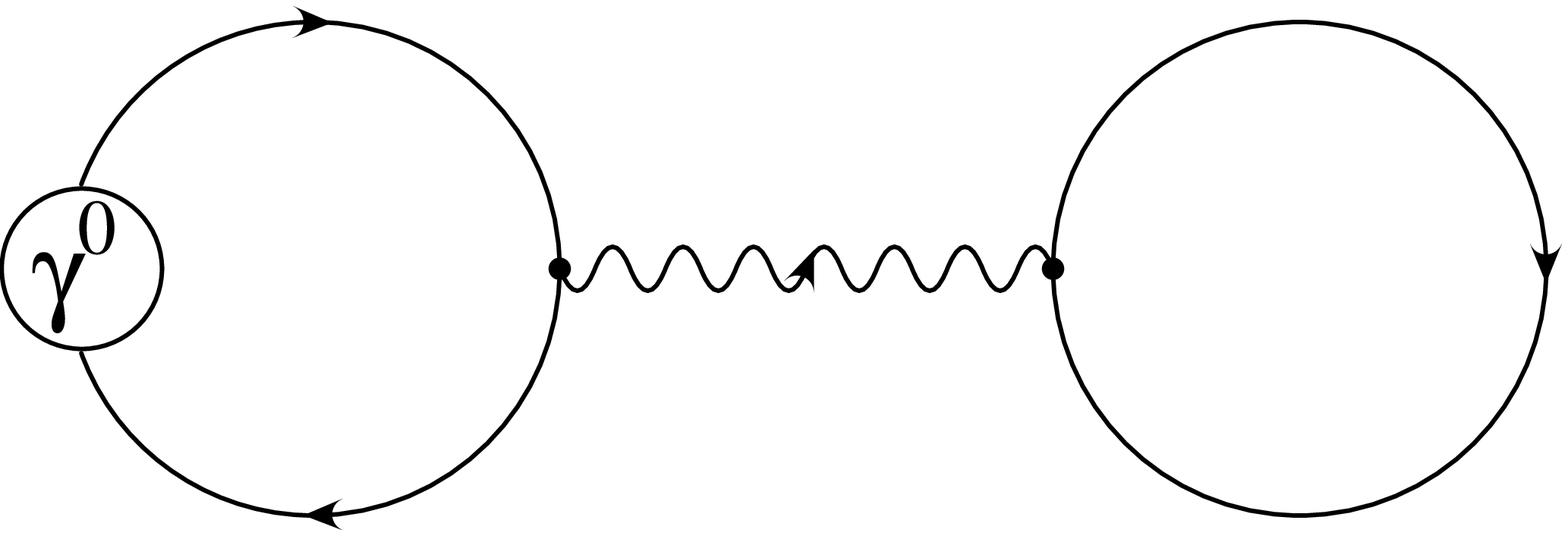,width=1.5cm}\end{array}
-\begin{array}{l}\psfig{file=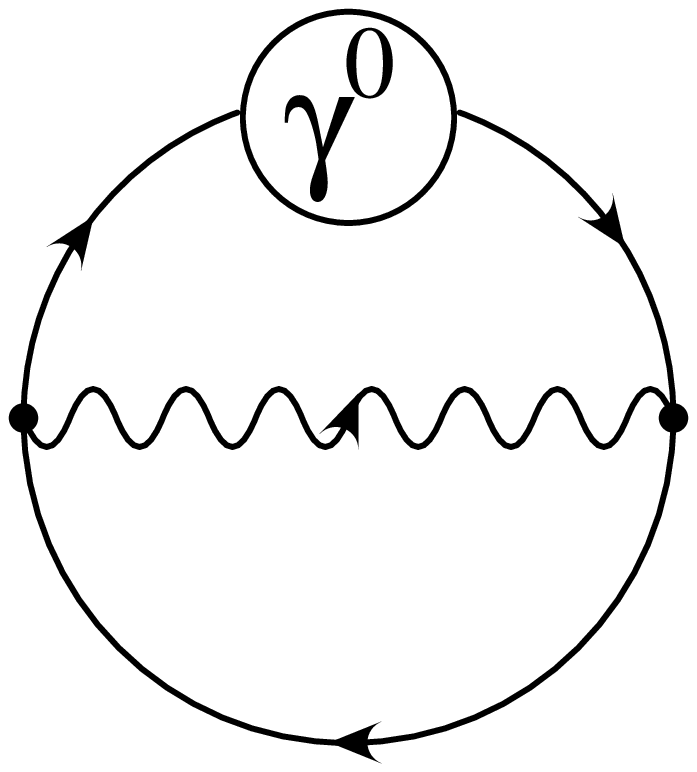,width=1cm}\end{array},
\label{2lcd}
\ee
for more details see App. \ref{app:cd}.
Charge neutrality for $\mu=0$ implies the renormalization condition
$\rho_\mr{ren}[\bar A\equiv 0]=0$ which can be satisfied by the subtraction
\be
\rho_\mr{ren}[\bar A]=\rho[\bar A]-\rho[0].
\label{rhosub}
\ee

\subsection{Numerical results}\label{phdia}
The energy densities given by Eqs.  \eq{1-2loop} and \eq{omcas}
have been calculated numerically for both the periodic and the
homogeneous phases using the  explicit formulae of App. \ref{2ldiags}
in units of $e=1$ and the details of the numerical procedure are
discussed in App. \ref{app:np}. The negative value of the Casimir energy,
found numerically, indicates that the periodic
state is energetically favoured. The  one- and two-loop contributions to the
energy are shown in Fig. \ref{1loop}. The one-loop  contribution, i.e.
that of the first diagram on the r.h.s. of Eq. \eq{1-2loop} is more important
than the two-loop terms in the parameter range studied.
The two-loop correction which is dominated  by the exchange diagrams,
the third and the fifth ones on the r.h.s of Eq. \eq{1-2loop}, tends to
destabilize the periodic state for $e\mu<m$ and to stabilize it for $e\mu>m$.
The periodic phase is stable mainly due to the gain arising from the sinking
of upper bands to negative energies and this gain is   taken into account
completely in our computation since the one-particle energy levels have been
calculated non-perturbatively. The photon exchange appearing at the two-loop
order becomes important in the region $e\mu\approx m$. The numerical results
for $e\mu/m \stackrel{>}{\sim} 1.12$ agree for the periodic and the homogeneous
ground states within the numerical accuracy.

The computation of the charge density yields the equation of state,
the relation between the energy and the charge density.
For each $\mu$ and $m$ we looked for the values of $a(\mu)$ and $Q(\mu)$ at
which the Casimir-energy density assumes its minimum. In this way
we obtained ${\cal E}_\mr{C}(a(\mu),Q(\mu),\mu)={\cal E}_\mr{C}(\mu)$,
and $\rho_\mr{ren}(\mu)$ for each fixed value of $m$.
The equation of state then can be constructed by tracing
${\cal E}_\mr{per}(\mu)$ and ${\cal E}_\mr{n}(\mu)$
as the function of $\rho_\mr{ren}(\mu)$ for the modulated and the homogeneous
phases, respectively. A typical $\rho$-dependence is shown in Fig.
\ref{rhores} for $m=2$, similar curves are found for the other values $m$.

The two-loop results shown in Fig. \ref{rhores}a indicate in a reliable
manner that the energy minimum for the periodic ground state,
${\cal E}_{\mr{per}}$, is smaller than that for the homogeneous ground
state, ${\cal E}_{\mr{n}}$ for $\rho\in[0,0.33]$. The amplitude
of the periodic background field tends to zero with increasing $\rho$
as one can see in Fig. \ref{rhores}b.
Our numerical results for $\rho \stackrel{<}{\sim} 0.33$ show
rapid oscillations in Figs. \ref{rhores}b and c reflecting a numerical
problem which arises due to the almost degeneracy found for
the periodic states with $Q=k_F$ and $2k_F$.
It was found that the wavenumber $Q$ of the periodic phase
is directly related to the Fermi momentum $k_F$ via the relation $f Q=2k_F$,
with the filling factor $f$, defined as the number of the entirely filled
positive-energy bands  plus the fraction of the partially filled band, see Fig.
\ref{brill}. The ratio $Q/k_F$ displays a discrete behaviour,
$Q/k_F\approx2$ for $e\mu<m$ and $Q/k_F\approx1$ for $e\mu>m$.
The discrete nature of $Q/k_F$ reveals that the vacuum
always readjusts itself until a forbidden zone is opened at the
Fermi-level and the filling factor becomes integer.
This observation is in agreement with the nesting relation $fQ=2 k_F$.
The product  $fQ$ shown in Fig. \ref{rhores}c increases in the
average with increasing charge density, implying the same behaviour
of the wavelength as was found in the bosonized theory,
c.f. Fig. \ref{g1logl}.

For densities larger than the 'critical density' $\rho>0.33$ we found no
energy difference between the periodic and the normal ground state within our
numerical accuracy. The reliable numerical determination of the parameters
$a$ and $Q$ became impossible, so that those are not shown in Figs.
\ref{rhores}b, c for large densities.
The result obtained in the framework of the bosonized model, namely the
small but non-vanishing amplitude of the modulation of the vacuum
for large densities as shown in Fig. \ref{g1loga} makes one cautious that
above the 'critical density' the numerical minimization fails to find
the proper minimum. But fortunately the smallness of the amplitude of
the induced field at large
densities enables one to treat the periodic potential as a perturbation.
The external potential \eq{ansatz} should open a gap around the Fermi level
at $k_F=Q/2$ according to the leading order of the degenerate perturbation
expansion \cite{Landau} and one finds the energy spectrum
\be
\epsilon^\pm_k=\frac12\left[\epsilon^{(0)}_k+\epsilon^{(0)}_{k-Q}
\pm\sqrt{(\epsilon^{(0)}_k-\epsilon^{(0)}_{k-Q})^2+4e^2a^2}\right],
\label{eig}
\ee
close to the quasi-momentum $k\approx k_F$,
where $\epsilon^{(0)}_k$ refers to the unperturbed spectrum and the signs
$+$ and $-$ stand for $|k|>\hf Q$ and $|k|<\hf Q$, respectively.
Due to the first order perturbation treatment we can now ignore the
two-loop order diagrams in Eq. (\ref{1-2loop}). By means of the energy
eigenvalues in Eq. (\ref{eig}) one finds
\be
{\cal E}_\mr{C}(a,Q,\mu)=\frac14 a^2Q^2+
\begin{array}{l}\psfig{file=OneLoopEnergy.eps,width=1cm}\end{array}_\mr{per}-
\begin{array}{l}\psfig{file=OneLoopEnergy.eps,width=1cm}\end{array}_\mr{n}
=\frac14a^2Q^2+
\int_{-Q/2}^{Q/2}\frac{\mr{d}k'}{2\pi}\epsilon^-_{\hf Q-k'}-
\int_{-Q/2}^{Q/2}\frac{\mr{d}k'}{2\pi}\epsilon^{(0)}_{\hf Q-k'},
\label{ecasf}
\ee
for the Casimir-energy density in the leading order.
Since $\rho=\int_{-k_F}^{k_F}~ dk/2\pi=k_F/\pi$ and
$Q/2=k_F=\pi\rho$ we have $\rho=Q/2\pi$.
For large values of the chemical potential $\mu$ we expect large values of
$k_F=Q/2\gg m$ therefore we can set $m=0$, i.e. $\epsilon^{(0)}(k)=|k|$,
$\mu=k_F=Q/2$ and one can perform the integral in Eq. (\ref{ecasf}) explicitly,
\be
{\cal E}_\mr{C}(a,Q,Q/2)=\frac{Q^2}{8\pi}+\frac14a^2Q^2
-{Q\sqrt{Q^2+4e^2a^2}\over 8\pi}-{e^2a^2\over 2\pi}\mr{arsinh}\frac{Q}{2ea}.
\label{ecasm0}
\ee
For a given $Q$, i.e. $\rho$ the Casimir-energy density has two
extrema: a maximum at
$a=0$ $(\partial_a^2{\cal E}_\mr{C}(0,Q,Q/2)\to-\infty)$ and a minimum at
\bea
a_\mr{min}=\frac{Q}{2e\sinh({\pi Q^2\over2e^2})}
=\frac{\pi\rho}{e\sinh({2\pi^3\rho^2\over e^2})}>0
\eea
$(\partial_a^2{\cal E}_\mr{C}(a_\mr{min},Q,Q/2)>0)$. The latter
provides the Casimir-energy density of the ground state,
\be
{\cal E}_\mr{C}(a_\mr{min},Q,Q/2)=\frac{Q^2}{8\pi}\left[1-
\sqrt{1+{4e^2a_\mr{min}^2\over Q^2}}\right]
=\frac{\pi\rho^2}{2}\left[1-\coth\frac{2\pi^3\rho^2}{e^2}\right]<0.
\ee
One finds $a_\mr{min}\approx 0.004$ and
${\cal E}_\mr{C}(a_\mr{min},2\pi\rho,\pi\rho)\approx -10^{-5}$
at the `critical density' $\rho=0.33$ therefore it is not possible to
confirm the periodicity of the ground state within our numerical
accuracy. Furthermore we have seen for lower densities that the jump
from one band to two bands in the Dirac sea is a two-loop effect
caused by photon exchange. In general, one needs higher-loop corrections
in order to let more than two bands sinking into the Dirac sea.
This is due to the observation that the $N$-th order perturbation
expansion in the monochromatic external field predicts the opening of
$N$ gaps in the fermion spectrum. Nevertheless this simple computation
indicates that the vacuum of the massive Schwinger model
keeps its periodicity, although with  decreasing amplitude
$a\sim \rho e^{-2\pi^3 \rho^2/e^2}$ and wavelength $2\pi/Q=1/\rho$
for increasing charge density $\rho$, in a manner similar to
the bosonized model (see Fig. \ref{g1logl}).
The one-loop level vacuum for the assumed simple sinusoidal
background potential involves a single band sunk into the Dirac sea.

For the low-density periodic phase
$e\mu\approx m$ where the amplitude $a$ is non-perturbative the photon
exchange is significant. As $\rho$ increases the perturbative region of $a$
is reached, although we could not decide, whether the numerically found
`critical density' does fall into the perturbative region. Our one-loop
perturbative result cannot clarify whether higher-order loop corrections
can  lead to sinking more and more bands into the Dirac sea as the density
$\rho$ increases.

\section{Summary}\label{summary}

The ground state for the massive Schwinger model has been investigated in the
presence of  homogeneous external charge density. The energy density of the
ground state has been determined numerically in the bosonized version of the
model as well as in terms of the original degrees of freedom of QED
by variational methods. The scalar field configuration and the charge
density of the ground state for the bosonized model have been obtained
by minimizing the tree-level energy in the presence of static,
homogeneous external charge density. In the fermionic theory a
variational method has been constructed by minimizing for the
amplitude and the wavelength of a sinusoidal photon condensate
in the vacuum. The finite charge density was realized by the introduction
of the chemical potential. The applicability of the loop-expansion
for the computation of the vacuum energy at finite density is justified
by the bosonized theory which indicates that the confining
Coulomb-force among integer charges is vanishing.

The computation in the bosonized theory shows that the system
exhibits a periodic ground state for arbitrary charge density.
The fermionic computation gives the same result. Numerical computations
reliable up to a certain charge density as well as simple analytic
calculations valid for asymptotically large charge densities support
the periodicity of the ground state. The general trends in the
charge-density dependence of the amplitude and that of the wavelength of
the periodic structure are in agreement for both versions of the model.

The investigations in terms of the bosonic and fermionic degrees of freedom
complement each other. On the one hand, our results for the bosonized model
showed that the background charge density is in average neutralized in
the ground state. Furthermore, the charge density wave ground state
and the complete screening of the integer charges appear
due to the interplay of the kinetic energy, the periodic potential
energy and that of the boundary condition
for the boson field. The fermionic description, on the other hand,
gives more insight into the structure
of the vacuum, namely that the modulation of the charge density
arises as the result of the opening of a gap in the fermion spectrum.
The conclusion of the straightforward perturbation expansion
for large charge densities is that the periodic ground state is
always favoured by the system energetically as compared to the homogeneous
one, even if only a single band sinks into the Dirac sea.
In the case of a single mode periodic potential $N$ bands sink below
the Fermi-level in the $N$-th order of the perturbation expansion.
In particular, in our numerical computation where a single mode
was allowed for the induced photon field in the vacuum and first
(second) order perturbation expression for the vacuum energy was
minimized one (two) bands are found below the Fermi-level for large
densities. The Peierls mechanism is present in the relativistic
vacuum for $N\ge2$. It is however not clear how to identify this
mechanism in terms of the bosonic excitations.

Our analytic considerations showing the existence of the periodic ground
state for arbitrarily large charge densities do not include the
corresponding higher-order loop corrections neither for the bosonized
model, nor for the fermionic one. It still remains an open question
whether a complete resummation of these higher-order corrections
would alter the qualitative result obtained here, namely that the
massive Schwinger model has only a single phase, the periodic one for any
values of the average charge density.

\section*{Acknowledgements}
This work has been supported by the grants
OTKA T032501/00, NATO SA(PST.CLG 975722)5066, and also
partially by the grant  OTKA M041537 and  the Supercomputing
Laboratory of the Faculty of Natural Sciences, University of Debrecen.
One of us (K.S.) has been supported by the Alexander von
Humboldt-Foundation and he thanks W. Greiner for the kind hospitality
and the impetus given by him to study strongly interacting vacua
and also thanks J. Csikai making him aware of Ref. \cite{Slater 1952}
in the general physics lectures nearly three decades ago.

\appendix

\section{Dirac equation with sinusoidal potential}\label{DH}
The Lehmann representation of the non-interacting electron propagator requires
the knowledge of the eigenfunctions of the Dirac Hamiltonian  \eq{dirham} for
the external field \eq{ansatz}. We  use the real Dirac matrices
\begin{equation}\label{gamma}
\gamma^0=
\left(\begin{array}{cc}
1 & 0\\
0 & -1\\
\end{array}\right) ,\qquad
\gamma^1=
\left(\begin{array}{cc}
0 & 1 \\
-1 & 0 \\
\end{array}\right).
\end{equation}
The  eigenspinors $f^{ks} (x)$ and $g^{ks} (x)$ of the Dirac Hamiltonian
$H_D ( {\bar A} )$ belonging to the positive and negative energy
eigenvalues $\epsilon_{ks}^{(+)}{>}0$ and $-\epsilon_{ks}^{(-)}{<}0$,
respectively satisfy the equations
\begin{eqnarray}\label{diracegy}
H_D ( {\bar A} ) \ f^{ks} (x) = \epsilon^{(+)}_{ks} \ f^{ks} (x),
~~~~
H_D ( {\bar A} ) \ g^{ks} (x) = -\epsilon^{(-)}_{ks} \ g^{ks} (x),
\label{H_D}
\end{eqnarray}
where the quasi-momentum $k \in \left[ - Q/2, Q/2 \right)$ takes values in the
first Brillouin-zone. The non-negative integer $s \ge 0$ labels the bands in
increasing order in the energy. As in the non-relativistic case, the solutions
of Eq. (\ref{H_D}) are Bloch-waves,
\begin{eqnarray}
f^{ks}=\sum_{n=-\infty}^{\infty}u^{ks}_n
e^{-{\rm i}\left(\epsilon_{ks}^{(+)} x_0-k_n x_1\right)},~~~~
g^{ks}=\sum_{n=-\infty}^{\infty}v^{ks}_n
e^{{\rm i}\left(\epsilon_{ks}^{(-)} x_0-k_n x_1\right)},
\label{f-g}
\end{eqnarray}
with $k_n{=}k{+}nQ$. In order to find the numerical solution, one rewrites
Eqs. \eq{diracegy} in matrix forms for the components of the Bloch-waves,
e.g. the first one of Eqs. \eq{diracegy} reads as
\begin{equation}
\sum_{n=-\infty}^\infty\left[\left(\epsilon_{ks}+e\mu-
k_n\gamma^0\gamma^1-m\gamma^0\right)u^{ks}_n+\frac{ea}2\left(
u^{ks}_{n+1}+u^{ks}_{n-1}\right)\right]e^{-{\rm i}(\epsilon_{ks} x_0-k_n x_1)}
=0.
\label{linseq}
\end{equation}
The solution is found by making up a matrix from the coefficients appearing
next to the Dirac spinors $u^{ks}_n$. The non-relativistic treatment
results in a matrix with tridiagonal structure \cite{Slater 1952},
\cite{G_R}. The structure of the matrix remains unchanged in the
relativistic case except of the replacement of the matrix elements with
$2\times 2$ matrices. The problem is then reduced to solve
a system of a coupled set of homogeneous linear equations.

\section{Band structure}\label{enspec}
In order to understand the effects of the photon exchanges,
the salient features of non-interacting electrons in static periodic
background field are briefly summarized in this section.
The eigenspinors and the energy eigenvalues of the Dirac
Hamiltonian \eq{dirham} in the static, periodic external field
have been determined numerically  (App. \ref{DH}).
The positive and negative single-particle energies
are denoted by  $\epsilon^{(+)}_{ks}$ and $-\epsilon^{(-)}_{ks}$, respectively,
as the functions of the quasi-momentum $k$ and the band index $s$.
Their dependences on the parameters $a$, $Q$  are not
indicated explicitly. Since the chemical potential $\mu$ results
in a constant shift of the whole fermion spectrum, it is sufficient to
understand the spectrum for $\mu=0$. Due to the periodic potential,
a band structure with  alternating allowed and forbidden bands
is formed \cite{Das 1973}. The typical band structure is plotted in Fig.
\ref{relslat}a and \ref{relslat}b as the function of $1/Q$ for
undercritical $ea<m$  and overcritical $ea>m$ static periodic
external electric fields, respectively. The shaded regions in Figs.
\ref{relslat} correspond to the allowed bands.
The non-relativistic analogue of these figures can be
found in \cite{Slater 1952} where the one-dimensional electron system was
considered in the presence of the static  external electric potential
$A_0=a(1-\cos Qx)$ (with our notations) that is just the same potential
we have but one of its minima is shifted to $x=0$ and the potential is chosen
zero in this minimum. The following qualitative features of the fermion
spectrum are wothwhile mentioning.
\begin{enumerate}
\item
Undercritical vacuum $ea{<}m$: The mass gap around zero energy separates
the infinite towers of bands above and below this gap. The smaller is
${1/ Q}$, the more the allowed bands widen out and start to overlap.
This is just the qualitative behaviour  obtained in the non-relativistic
description \cite{Slater 1952}. The energies of the states in the
upper, $\epsilon>0$ (lower, $\epsilon<0$) tower decrease (increase) with
increasing $1/Q$.
\item Overcritical vacuum $ea>m$: The qualitative features described above
remain the same but the upper and the lower bands overlap for large enough
$1/Q$ and upper bands sink into the Dirac sea while lower bands emerge.
The band crossing is well understood \cite{Nagy 2 2002} but this case
which involvs the creation of electron-positron pairs turned out
to be not relevant for our purpose because the periodic ground state
was found undercritical.
\end{enumerate}

The asymptotics of the spectrum for $1/Q\to0$ and $1/Q\to\infty$
is helpful to understand the $1/Q$-dependence of the band structure.
\begin{enumerate}
\item $1/Q\to 0$: The energy levels decouple from the periodic structure
of the potential and the free fermion spectrum reappears
with the single gap for $-m\le\epsilon\le m$. This can easily understood
by noticing that the potential term in the Dirac Hamiltonian \eq{dirham}
becomes negligible compared to the kinetic energy for $Q \to \infty$. In fact,
the introduction of the rescaled coordinate $\xi_1=Q x_1$ leaves the
only $Q$-dependence coming from the gradient term of the Hamiltonian.
This corresponds to infinitely densely packed atoms in the model of
Ref. \cite{Slater 1952} and to vanishing average electric potential.
\item $1/Q\to\infty$: The extrema of the potential is now well-separated
and one expects localized states at the minima (maxima)
corresponding to the upper (lower) bands. Furthermore each band
should be reduced to a single, highly degenerate energy level
which corresponds to the localized states at the various minima
(maxima) of the external potential.
The semiclassical tunneling probability from a minimum to the neighbouring
one is suppressed exponentially with increasing $1/Q$,
$w\sim \exp\{-16\sqrt{mea}/Q\}$ and  the problem reduces to that of the
relativistic harmonic oscillator as far as the lowest (highest) lying states
of the upper (lower) band are concerned \cite{Nagy 2 2002}.
\end{enumerate}

The dispersion relations in the allowed bands  alternate between convex
and concave ones from band to band. If the Fermi-level lies inside of an
allowed band, the occupied states build either a Fermi sphere or a Fermi hole
in momentum space. For one spatial dimension the Fermi sphere is distorted
to a Fermi section $p_1 \in [-k_F,k_F]$, the Fermi hole appears as the
unoccupied interval $p_1 \in [-k_F,k_F]$ between the occupied ones,
$p_1\in [-Q/2,-k_F]$ and  $p_1\in [k_F, Q/2]$.

\section{Numerical calculation of the energy and charge density}\label{app:en}

This appendix contains the details of the calculations of the energy and charge
density.

\subsection{Bare expression for the energy density}\label{2lform}
The effective Lagrangian corresponding to the action \eq{genfdf},
\be
{\cal L}=-\frac14F_{\rho\sigma}F^{\rho\sigma}
+\hf\psib i\gamma^\rho\partial_\rho\psi-\hf\partial_\rho\psib i\gamma^\rho\psi
+e\psib\gamma^\rho A_\rho\psi-m\psib\psi
+\lambda\frac14(F-\bar F)_{\rho\sigma}\bar F^{\rho\sigma}
\ee
yields the energy-momentum tensor,
\be
T^{\mu\nu}=\frac{\partial{\cal L}}{\partial(\partial_\mu A^\kappa)}
\partial^\nu A^\kappa-\frac{\partial{\cal L}}{\partial(\partial_\mu\psi)}
\partial^\nu\psi+\partial^\mu\psib\frac{\partial{\cal L}}
{\partial(\partial_\nu\psib)}+\lambda\bar F^{\mu\kappa}\partial^\nu A_\kappa
-g^{\mu\nu}{\cal L},
\ee
which should be symmetrized by adding the
divergence $-\partial_\rho f^{\nu\rho\mu}$ of the third rank tensor
\be
f^{\nu\mu\rho}=F^{\mu\rho}A^\nu+\frac{\ci}8\psib(\gamma^\mu\gamma^\nu
\gamma^\rho-\gamma^\rho\gamma^\nu\gamma^\mu)\psi,
\ee
determined by the spin density,
\be
T^{\mu\nu}_\mr{sym}=T^{\mu\nu}-\partial_\rho f^{\nu\rho\mu}
=-F^{\mu\rho}F^\nu_\rho+\lambda\bar F^{\mu\kappa}\partial^\nu A_\kappa
+\frac{\ci}4(\psib\gamma^\mu\partial^\nu\psi+\psib\gamma^\nu\partial^\mu\psi)
-\frac{\ci}4(\partial^\nu\psib\gamma^\mu\psi
+\partial^\mu\psib\gamma^\nu\psi)-g^{\mu\nu}{\cal L}.
\ee
It is easy to see that the symmetrized energy-momentum tensor is gauge
invariant. The energy density operator is
\be
\hat {\cal E}[A,\psib,\psi]=\frac1{LT}\int_xT^{00}_\mr{sym}
=\frac1{LT}\int_x\left[\hf(F^{01}_x)^2-\psib_xH_D(A)\psi_x
+\lambda\bar F^{0\kappa}\partial^0 A_\kappa
-\frac{\lambda}4(F-\bar F)_{\rho\sigma}\bar F^{\rho\sigma}\right],
\ee
where the first term on the r.h.s. is the energy density of the photons and
the second one is the energy density of the Dirac sea minus $e\mu$
multiplied by the fermion density. The last two terms appear due to
the exclusion of the fluctuations of the collective mode. One can write
$\hat{\cal E}=\hat{\cal E}_{(0)}+\hat{\cal E}_{(1)}+\hat{\cal E}_{(2)}$ with
\bea\label{gseloop}
\hat{\cal E}_{(0)}[\bar A,\psib,\psi,\lambda]&=&
\frac1{LT}\int_x\left[\hf(\bar F^{01}_x)^2
+\lambda\bar F^{0\kappa}\partial^0\bar A_\kappa
-\psib_xH_D(\bar A)\psi_x\right],\nonu
\hat{\cal E}_{(1)}[\bar A,\alpha,\psib,\psi,\lambda]&=&\frac1{LT}\int_x
\alpha_{\mu x}\left(-g^{\mu1}\partial_x^0\bar F^{01}_x
+g^{\mu0}\partial_x^1\bar F^{01}+\frac{\lambda}2\Box_x\bar A_x^\mu
-\lambda\partial_x^0\bar F^{0\mu}_x-e\psib\gamma^\mu\psi\right),\nonu
\hat{\cal E}_{(2)}[\alpha]&=&\frac1{LT}\int_x\hf
[(g^{\mu 1}\partial^0-g^{\mu0}\partial^1)\alpha_{\mu x}]
[(g^{\nu1}\partial^0-g^{\nu0}\partial^1)\alpha_{\nu x}],
\eea
where the lower indices indicate the powers of $\alpha$. According to
Eq. \ref{expvalO}, the expectation
value of the operator $\hat{\cal E}$ in the vacuum is given as
${\cal E}[\bar A]=\sum_{i=0}^2{\cal E}_{(i)}[\bar A]$. We find
\bea\label{opzero}
{\cal E}_{(0)}[\bar A]&=&\frac14Q^2a^2+\frac1{Z}
\int_x \frac{\delta}{{\rm i}\delta\zeta_\alpha^x}
\hat H_D(\bar A)_{\alpha\beta}\frac{\delta}{{\rm i}\delta\bar\zeta_\beta^x}
Z_0|_{j=\zeta=\bar\zeta=0}\nonu
&&-\frac{e^2}{2Z}\int_x \frac{\delta}{{\rm i}\delta\zeta_\alpha^x}
\hat H_D(\bar A)_{\alpha\beta}\frac{\delta}{{\rm i}\delta\bar\zeta_\beta^x}
\int_{y,z}\fd{}{\zeta_\kappa^y}\gamma^\mu_{\kappa\lambda}\fd{}{j_\mu^y}
\fd{}{\bar\zeta_\lambda^y}\fd{}{\zeta_\epsilon^z}\gamma^\nu_{\epsilon\delta}
\fd{}{j_\nu^z}\fd{}{\bar\zeta_\delta^z}Z_0|_{j=\zeta=\bar\zeta=0}\nonu
&=&\frac14Q^2a^2+\frac{1}{Z}\left[
\begin{array}{l}\psfig{file=OneLoopEnergy.eps,width=1.1cm}\end{array}
-\hf\begin{array}{l}\psfig{file=OneLoopEnergy.eps,width=1.1cm}\end{array}
\begin{array}{l}\psfig{file=Glasses.eps,width=1.9cm}\end{array}
+\begin{array}{l}\psfig{file=GlassesIncH.eps,width=1.9cm}\end{array}
+\hf\begin{array}{l}\psfig{file=OneLoopEnergy.eps,width=1.1cm}\end{array}
\begin{array}{l}\psfig{file=Theta.eps,width=1.1cm}\end{array}
-\begin{array}{l}\psfig{file=ThetaIncH.eps,width=1.1cm}\end{array}\right]
\eea
with
\be
Z_0=\exp\left\{-\frac{\ci}2j\cdot D\cdot j-\ci\bar\zeta\cdot G(\bar A)\cdot
\zeta\right\},
\ee
and $\hat H_D(\bar A)$ denoting the Dirac Hamiltonian (\ref{dirham}) with
the field variables replaced by functional derivatives with respect to the
corresponding external sources, and the vacuum-to-vacuum amplitude up
to the order $\ord{e^2}$ is
\bea
Z&=&1-\frac{e^2}2\int_{x,y}\fd{}{\zeta_\alpha^x}\gamma^\mu_{\alpha\beta}
\fd{}{j_\mu^x}
\fd{}{\bar\zeta_\beta^x}\fd{}{\zeta_\epsilon^y}\gamma^\nu_{\epsilon\delta}
\fd{}{j_\nu^y}
\fd{}{\bar\zeta_\delta^y} Z_0|_{j,\zeta,\bar\zeta=0}\nonu
&=&1-\frac12\left[
\begin{array}{l}\psfig{file=Glasses.eps,width=2cm}\end{array}-
\begin{array}{l}\psfig{file=Theta.eps,width=1.2cm}\end{array}\right].
\eea
The $\ord{\alpha^0}$ energy, Eq. \eq {opzero}, includes the energy density
of the background and those of the modulated, interacting Dirac-sea.
The $\ord{\alpha}$ energy term is the interaction energy
of the current with the fluctuations of the photon field,
\bea\label{opone}
{\cal E}_{(1)}[\bar A]&=&-\frac{\ci e^2}{Z}\int_{x,y}
\frac{\delta}{{\rm i}\delta\eta_\alpha^x}\gamma^\mu_{\alpha\beta}
\frac{\delta}{{\rm i}\delta j_\mu^x}\frac{\delta}{{\rm i}
\delta\bar\eta_\beta^x}
\frac{\delta}{{\ci}\delta\eta_\alpha^y}\gamma^\nu_{\alpha\beta}
\frac{\delta}{{\ci}\delta j_\nu^y}
\frac{\delta}{{\ci}\delta\bar\eta_\beta^y}Z_0|_{j=\eta=\bar\eta=0}\nonu
&=&-\frac{\ci}{Z}\left[
\begin{array}{l}\psfig{file=Glasses.eps,width=2cm}\end{array}
-\begin{array}{l}\psfig{file=Theta.eps,width=1.2cm}\end{array}\right].
\eea
The $\ord{\alpha^2}$ energy expression depends on the fluctuating
field $\alpha$ only.
Since the photon propagator in the presence of the background field
approaches the free propagator as $L\to\infty$ the contribution
${\cal E}_{(2)}$ cancels when the difference of the energy densities with
and without the background field is considered.
The vacuum to vacuum amplitude removes the disconnected components as
expected and one finds
\be
{\cal E}[\bar A]= \frac14 a^2Q^2+
\begin{array}{l}\psfig{file=OneLoopEnergy.eps,width=1cm}\end{array}
-\ci\begin{array}{l}\psfig{file=Glasses.eps,width=1.5cm}\end{array}
+\ci\begin{array}{l}\psfig{file=Theta.eps,width=1cm}\end{array}
+\begin{array}{l}\psfig{file=GlassesIncH.eps,width=1.5cm}\end{array}
-\begin{array}{l}\psfig{file=ThetaIncH.eps,width=1cm}\end{array}
\label{app:1-2loop}.
\ee
The propagators are calculated by means of the Lehmann representation
\be
G^{\alpha\beta}_{xy}=
\sum_{k_1s_1}\int \frac{dk_0}{2\pi}e^{-\ci k_0(x_0-y_0)}
\Bigl[\frac{f^{k_1s_1}_\alpha(x)\bar f^{k_1s_1}_\beta(y)}
{k_0+i\varepsilon}+\frac{g^{k_1s_1}_\alpha(x)\bar g^{k_1s_1}_\beta(y)}
{k_0-i\varepsilon}\Bigr],~~~~
D^{\mu\nu}_{xy} = -g_{\mu\nu}\sum_{k_1}\int \frac{dk_0}{2\pi}
\frac1{k^2+i\varepsilon} e^{\ci k(x-y)},
\ee
where $f^{k_1s_1}(x)$ and $g^{k_1s_1}(x)$ denote the positive and negative
energy eigensolutions of the Dirac-equation (see App. \ref{DH}).
The periodic background potential breaks the translational symmetry
which manifests itself in changing momentum conservation to
quasi-momentum conservation in each vertex.

\subsection{UV and IR divergences}\label{UVIR}
The first diagram on the r.h.s. of Eq. \eq{app:1-2loop} represents
the energy of the Dirac-sea in the presence of the background field,
\be\label{oneloopen}
{\cal E}_\mr{sea}(a,Q,\mu)=
\begin{array}{l}\psfig{file=OneLoopEnergy.eps,width=1.0cm}\end{array}
={1\over LT}\int dx\gamma^0_{\alpha\beta}\hat H_D^{x(\beta)}(\bar A)
\ci G_{xx}^{\beta\alpha}
=-{1\over LT}\sum_{k_1s_1}\epsilon^{(-)}_{k_1s_1},
\ee
and is quadratically divergent in the absence of the background field,
\be
{\cal E}_\mr{sea}(0,0,0)=-\int_{-\Lambda}^\Lambda \frac{\rm{d}p}{2\pi}
\left[p^2+m^2\right]^{1/2}.
\ee
The finite, physical part of ${\cal E}_\mr{sea}$ will be defined by
\be
{\cal E}_\mr{per}^{(1-l)}(a,Q,\mu)=
{\cal E}_\mr{sea}(a,Q,\mu)-{\cal E}_\mr{sea}(0,0,0),
~~~~
{\cal E}_\mr{n}^{(1-l)}(\mu)={\cal E}_\mr{sea}(0,0,\mu)-
{\cal E}_\mr{sea}(0,0,0)
\ee
for the periodic and the homogeneous, normal phases, respectively.

The convergence of ${\cal E}_\mr{per}$ was checked numerically in the
following manner. The one-loop contributions are obtained by taking
$\cal E$ of Eq. \eq{oneloopen} for the background
field and subtracting from it the same diagram without background field.
Let us consider first this difference for vanishing chemical potential $\mu=0$,
\be
{\cal E}_\mr{per}^{(1-l)}(a,Q,0)=-{1\over LT}\sum_{k_1\;s_1}
[\epsilon^{(-)}_{k_1\;s_1}(a,Q,0)-\epsilon^{(-)}_{k_1\;s_1}(0,0,0)].
\label{1l-sub}
\ee
By the one-by-one identification of the corresponding levels
we found numerically that the magnitude
$|\epsilon_{k_1\;s_1}(a,Q,0)-\epsilon_{k_1\;s_1}(0,0,0)|$
is suppressed for increasing $k_1$ according to the power law
\be
|\epsilon^{(-)}_{k_1\;s_1}(a,Q,0)-\epsilon^{(-)}_{k_1\;s_1}(0,0,0)|
\approx k_1^{-(1+\delta)}
\ee
with $\delta\approx2>0$, cf. Fig. \ref{konv}. This renders the sum
absolutely convergent. The shift in the spectrum caused by the non-vanishing
chemical potential does not alter the UV behaviour of the sum even in the
thermodynamic limit. The convergence of
${\cal E}_\mr{n}(\mu)={\cal E}(0,0,\mu)-{\cal E}(0,0,0)$ has been checked
similarly.

IR divergences can also appear at the tadpoles where the photon line
carries vanishing momentum $q^\mu=0$. Since there is actually no dynamical
photon-field variable with vanishing energy and momenta, such tadpoles pose
no problem in the homogeneous, normal vacuum \cite{Kapusta,Toimela,Freedman}.
In the  periodic vacuum the photon can borrow the momentum $nQ$ from the
vacuum by the summation for $n\not=0$ and the tadpoles are finite.

\subsection{Charge density}\label{app:cd}

In order to understand the structure of the vacuum, we  need another
important observable, the average charge density $\rho$ given as the two-loop
order expectation value of the operator
\be
\hat\rho=\frac1{LT}\int_x\psib\gamma^0\psi.
\ee
The expectation value of $\rho[\bar A]$ is taken by Eq. \eq{expvalO}, and
truncated at the two-loop order is given as
\be
\rho[\bar A]=\begin{array}{l}\psfig{file=OneLoopG0.eps,width=1cm}
\end{array}
+\begin{array}{l}\psfig{file=GlassesIncG0.eps,width=1.5cm}\end{array}
-\begin{array}{l}\psfig{file=ThetaIncG0.eps,width=1cm}\end{array} .
\label{app:2lcd}
\ee
These diagrams are similar to the first, fourth and fifth ones
of Eq. \eq{app:1-2loop} except that the Hamiltonian-insertion is replaced by a
$\gamma^0$-insertion.
The renormalization prescription (\ref{rhosub})
removes the UV divergence of the first
diagram in Eq. (\ref{app:2lcd}), too. The second diagram gives vanishing
contribution for vanishing periodic background electric field
due to Furry's theorem. The calculation of these diagrams proceeds
like those for the Casimir-energy density.

\subsection{Two-loop diagrams}\label{2ldiags}
We present now the explicit expressions for the two-loop diagrams
on the r.h.s. of Eq. \eq{1-2loop}. The eigenspinors $u$ and $v$ are
defined in Eq. \eq{f-g} and the diagrams containing tadpoles are
\bea
\begin{array}{l}\psfig{file=Glasses.eps,width=2cm}\end{array}
&=&-\frac{\ci e^2}{LT}\int_{x,y}\ci D^{yx}_{\nu\mu}\gamma^\nu_{\alpha\beta}
\ci G^{xx}_{\beta\alpha}\gamma^\mu_{\epsilon\delta}
\ci G^{yy}_{\delta\epsilon}\nonu
&=&\frac{e^2}{LT}\sum_{q_1}\frac1{|q_1|^2}
\mathop{\mathop{\sum}_{k_1,s_1}}_{n_1,n_2}
\bar v_{\alpha n_1}^{k_1s_1}\gamma^\mu_{\alpha\beta}
v_{\beta n_2}^{k_1s_1}\delta(q_1+(n_1-n_2)Q)
\mathop{\mathop{\sum}_{k_1,s_1}}_{n_1,n_2}
\bar v_{\epsilon n_1}^{k_1s_1}\gamma^\mu_{\epsilon\delta}
v_{\delta n_2}^{k_1s_1}\delta(-q_1+(n_1-n_2)Q) ,
\end{eqnarray}
\begin{eqnarray}
\begin{array}{l}\psfig{file=GlassesIncH.eps,width=2cm}\end{array}
&=&\frac{e^2}{LT}\int_{x,y,z}\hat H^{x(\beta)}_D
  \ci D^{zy}_{\mu\nu}\gamma^0_{\alpha\beta}
  \ci G^{xy}_{\beta\kappa} \gamma^\mu_{\kappa\lambda}
  \ci G^{yx}_{\lambda\alpha} \gamma^\nu_{\epsilon\delta}
  \ci G^{zz}_{\delta\epsilon}\nonu
&& ~~ =
\frac{e^2}{LT}
\sum_{q_1}\frac1{|q_1|^2}
\mathop{\mathop{\sum}_{k_1,s_1}}_{n_1,n_2}
\bar v^{k_1s_1}_{\epsilon n_1}\gamma^\mu_{\epsilon\delta}
      v^{k_1s_1}_{\delta n_2}\delta_{q_1-Q(n_1-n_2)}\nonu
&& ~~
\times\Biggl\{
\mathop{\mathop{\sum}_{k_1,p_1}}_{s_1,s_2}
\frac{\sum_{n_1,n_2}\bar u^{k_1s_1}_{\kappa n_1}\gamma^\mu_{\kappa\lambda}
      v^{p_1s_2}_{\lambda n_2}\delta_{p_{n_2}+k_{n_1}-q_1}
\sum_{n_1,n_2}\bar v^{p_1s_2}_{\alpha n_1}\gamma^0_{\alpha\beta}
    u^{k_1s_1}_{\beta n_2}
\epsilon^{(+)}_{k_1s_1}\delta_{p_{n_1}+k_{n_2}}}
{\epsilon^{(-)}_{p_1s_2}+\epsilon^{(+)}_{k_1s_1}}
\nonu
&& ~~
-\mathop{\mathop{\sum}_{k_1,p_1}}_{s_1,s_2}
  \frac{\sum_{n_1,n_2}
  \bar v^{k_1s_1}_{\kappa n_1}\gamma^\mu_{\kappa\lambda}
     u^{p_1s_2}_{\lambda n_2}\delta_{p_{n_2}+k_{n_1}+q_1}
\sum_{n_1,n_2}  \bar u^{p_1s_2}_{\alpha n_1}\gamma^0_{\alpha\beta}
  v^{k_1s_1}_{\beta n_2}
\epsilon^{(-)}_{k_1s_1}\delta_{p_{n_1}+k_{n_2}}}
{\epsilon^{(+)}_{p_1s_2}+\epsilon^{(-)}_{k_1s_1}}
\Biggr\}.\nonu
\eea
The exchange diagrams are given as
\begin{eqnarray}
\begin{array}{l}\psfig{file=Theta.eps,width=2cm}\end{array}
        &=& \frac{\ci e^2}{LT}
	\gamma^\nu_{\alpha\beta}\gamma^\mu_{\kappa\lambda}
                \int dx\int dy \ci G_{xy}^{\beta\kappa}
                \ci G^{\lambda\alpha}_{yx} \ci D^{\mu \nu}_{yx}
        \nonu
        &=&
        \frac{e^2}{2LT}
        \mathop{\mathop{\sum}_{k_1p_1q_1}}_{s_1s_2}
        \mathop{\mathop{\sum}_{n_1n_2}}_{n_3n_4}
         \Biggl[\frac{\bar v^{p_1s_2}_{\alpha n_4}
        \gamma^\mu_{\alpha\beta}u^{k_1s_1}_{\beta n_1}
        \delta_{k_{1,n_1}+p_{1,n_4}+q_1}
        \bar u^{k_1s_1}_{\kappa n_2}\gamma^\mu_{\kappa\lambda}
                v^{p_1s_2}_{\lambda n_3}
        \delta_{k_{1,n_2}+p_{1,n_3}+q_1}}
        {|q_1|(|q_1|+\epsilon^{(+)}_{k_1s_1}+\epsilon^{(-)}_{p_1s_2})}\nonu
        && +
        \frac{\bar u^{p_1s_2}_{\alpha n_4}
        \gamma^\mu_{\alpha\beta}v^{k_1s_1}_{\beta n_1}
        \delta_{k_{1,n_1}+p_{1,n_4}-q_1}
        \bar v^{k_1s_1}_{\kappa n_2}
        \gamma_{\mu\kappa\lambda}u^{p_1s_2}_{\lambda n_3}
                \delta_{k_{1,n_2}+p_{1,n_3}-q_1}}
        {|q_1|(|q_1|+\epsilon^{(-)}_{k_1s_1}+\epsilon^{(+)}_{p_1s_2})}\Biggr]
,\nonu
\eea
and
\bea
\begin{array}{l}\psfig{file=ThetaIncH.eps,width=2cm}\end{array}
        &=& - \frac{e^2}{2LT}\int_{x,y,z}
                \gamma^0_{\alpha\beta}\gamma^\mu_{\epsilon\delta}
                \gamma^\nu_{\kappa\lambda}\hat H_{D,x}^{(\beta)}(\bar A)(
                \ci D_{yz}^{\mu\nu} \ci G^{\lambda\epsilon}_{zy}
                \ci G^{\beta\kappa}_{xz}  \ci G^{\delta\alpha}_{yx})
        \nonu
&& ~~~=  \frac{e^2}{4LT}
        \mathop{\mathop{\sum}_{k_1p_1r_1q_1}}_{s_1s_2s_3}\frac1{|q_1|}
        \nonu
&&  ~~~    \Bigl[
        \frac{
        \sum_{n_1n_2}\bar u^{k_1s_1}_{n_1}\gamma_\mu v^{r_1s_3}_{n_2}
        \delta_{k_{1,n_1}+r_{1,n_2}+q_1}
        \sum_{n_1n_2}\bar u^{p_1s_2}_{n_1}\gamma^\mu u^{k_1s_1}_{n_2}
        \delta_{k_{1,n_2}-p_{1,n_1}+q_1}}
        {(|q_1|+\epsilon^{(+)}_{k_1s_1}+\epsilon^{(-)}_{r_1s_3})
        (\epsilon^{(+)}_{p_1s_2}+\epsilon^{(-)}_{r_1s_3})}
        \nonu
&&~~~
        \sum_{n_1n_2}\bar v^{r_1s_3}_{n_1}\gamma^0 u^{p_1s_2}_{n_2}
        [\epsilon^{(+)}_{p_1s_2}\delta_{p_{1,n_2}+r_{1,n_1}}]+
        \nonu
&&~~~
        \frac{\sum_{n_1n_2}
        \bar u^{k_1s_1}_{n_1}\gamma_\mu u^{r_1s_3}_{n_2}
        \delta_{k_{1,n_1}-r_{1,n_2}+q_1}
        \sum_{n_1n_2}\bar v^{p_1s_2}_{n_1}\gamma^\mu u^{k_1s_1}_{n_2}
        \delta_{k_{1,n_2}+p_{1,n_1}+q_1}}
        {(|q_1|+\epsilon^{(+)}_{k_1s_1}+\epsilon^{(-)}_{p_1s_2})
        (\epsilon^{(-)}_{p_1s_2}+\epsilon^{(+)}_{r_1s_3})}
        \nonu
&&~~~
        \sum_{n_1n_2}\bar u^{r_1s_3}_{n_1}\gamma^0 v^{p_1s_2}_{n_2}
        [-\epsilon^{(-)}_{p_1s_2}\delta_{p_{1,n_2}+r_{1,n_1}}]-
        \nonu
&&~~~      \frac{\sum_{n_1n_2}
        \bar u^{k_1s_1}_{n_1}\gamma_\mu v^{r_1s_3}_{n_2}
        \delta_{k_{1,n_1}+r_{1,n_2}+q_1}
        \sum_{n_1n_2}\bar v^{p_1s_2}_{n_1}\gamma^\mu u^{k_1s_1}_{n_2}
        \delta_{k_{1,n_2}+p_{1,n_1}+q_1}}
        {(|q_1|+\epsilon^{(+)}_{k_1s_1}+\epsilon^{(-)}_{p_1s_2})
        (|q_1|+\epsilon^{(+)}_{k_1s_1}+\epsilon^{(-)}_{r_1s_3})}
        \nonu
&&~~~
        \sum_{n_1n_2}\bar v^{r_1s_3}_{n_1}\gamma^0 v^{p_1s_2}_{n_2}
        [-\epsilon^{(-)}_{p_1s_2}\delta_{p_{1,n_2}-r_{1,n_1}}]+
        \nonu
&&~~~
        \frac{\sum_{n_1n_2}
        \bar v^{k_1s_1}_{n_1}\gamma_\mu u^{r_1s_3}_{n_2}
        \delta_{k_{1,n_1}+r_{1,n_2}-q_1}
        \sum_{n_1n_2}\bar u^{p_1s_2}_{n_1}\gamma^\mu v^{k_1s_1}_{n_2}
        \delta_{k_{1,n_2}+p_{1,n_1}-q_1}}
        {(|q_1|+\epsilon^{(-)}_{k_1s_1}+\epsilon^{(+)}_{p_1s_2})
        (|q_1|+\epsilon^{(-)}_{k_1s_1}+\epsilon^{(+)}_{r_1s_3})}
        \nonu
&&~~~
        \sum_{n_1n_2}\bar u^{r_1s_3}_{n_1}\gamma^0 u^{p_1s_2}_{n_2}
        [\epsilon^{(+)}_{p_1s_2}\delta_{p_{1,n_2}-r_{1,n_1}}]-
        \nonu
&&~~~
        \frac{\sum_{n_1n_2}
        \bar v^{k_1s_1}_{n_1}\gamma_\mu u^{r_1s_3}_{n_2}
        \delta_{k_{1,n_1}+r_{1,n_2}-q_1}
        \sum_{n_1n_2}\bar v^{p_1s_2}_{n_1}\gamma^\mu v^{k_1s_1}_{n_2}
        \delta_{k_{1,n_2}-p_{1,n_1}-q_1}}
        {(|q_1|+\epsilon^{(-)}_{k_1s_1}+\epsilon^{(+)}_{r_1s_3})
        (\epsilon^{(-)}_{p_1s_2}+\epsilon^{(+)}_{r_1s_3})}
        \nonu
&&~~~
        \sum_{n_1n_2}\bar u^{r_1s_3}_{n_1}\gamma^0 v^{p_1s_2}_{n_2}
        [-\epsilon^{(-)}_{p_1s_2}\delta_{p_{1,n_2}+r_{1,n_1}}]-
        \nonu
&&~~~
        \frac{\sum_{n_1n_2}
        \bar v^{k_1s_1}_{n_1}\gamma_\mu v^{r_1s_3}_{n_2}
        \delta_{k_{1,n_1}-r_{1,n_2}-q_1}
        \sum_{n_1n_2}\bar u^{p_1s_2}_{n_1}\gamma^\mu v^{k_1s_1}_{n_2}
        \delta_{k_{1,n_2}+p_{1,n_1}-q_1}}
        {(|q_1|+\epsilon^{(-)}_{k_1s_1}+\epsilon^{(+)}_{p_1s_2})
        (\epsilon^{(+)}_{p_1s_2}+\epsilon^{(-)}_{r_1s_3})}
        \nonu
&&~~~
        \sum_{n_1n_2}\bar v^{r_1s_3}_{n_1}\gamma^0 u^{p_1s_2}_{n_2}
        [\epsilon^{(+)}_{p_1s_2}\delta_{p_{1,n_2}+r_{1,n_1}}]\Bigr].
\end{eqnarray}

\subsection{Numerical procedure}\label{app:np}
The one-particle energy levels and spinors needed for the
calculation were determined by solving numerically
the system of linear equations \eq{linseq} and the sum over
the components of the Bloch-waves was truncated for $|n|\le25$.
This procedure provided us 50 one-particle energy levels and
spinors for each momentum in the first Brillouin zone.
It was tested on the fermion spectrum without background field that
such a truncation starts to cause noticeable error on the spectrum for
the band index $s\approx n$ when the bands are numerated in energetically
increasing order. Therefore, bands with $s\le20$ have been taken into account
in the calculations of the two-loop diagrams. Such a truncation
allowed us to detect the effects of the background field with
sufficient accuracy because it was found that for $ea\le m$
the distortion of the dispersion relation due to the background field
is only significant for states belonging to the bands in the vicinity of
$m$. For about 5 bands away from $m$ the deviation of the energy
levels with and without the periodic background field turns practically
to zero. The one- and two-loop diagrams of
Eq. (\ref{1-2loop}) were computed in the first Brillouin zone
$k_1\in[-Q/2,Q/2)$ at 40 and 10 points, respectively. The
calculation of the one-loop diagram required higher numerical accuracy
due to the numerical elimination of the UV divergence. We also made a
test calculation for 20 division points which corresponded
to a larger volume $L$ and found that the numerical accuracy is
about 10\% for the two-loop contribution to the Casimir-energy
density in the whole range of the parameter values.

The amplitude $a$ was chosen through several orders of magnitude from
$ea=10^{-4}$ corresponding to the perturbative regime
to $ae \approx m$ for which pair-production might occur.
The wavenumber of the background field was restricted to be
$Q=0.8,1,1.5$ in the computation. The significantly smaller values
are uninteresting from the point of view of the periodic state since the
electrons become well localized in the limit $1/Q\to\infty$
and it would cost too much energy to delocalize them.
The other limit $1/Q\to 0$ is computationally time-consuming
since one has to take more Brillouin-zones into account.
Therefore the summations over the band index $s$, as well as over the
one-particle state's index $n$ have to be truncated at increasingly
higher values in the formulae of App. \ref{2lform}. Consequences of this very
restricted search of the minimum of the Casimir-energy density not allowing
for the higher values $Q >1.5$ make our numerical results unreliable for large
densities. The calculations were performed on an AlphaServer DS20 500 MHz
with 2 CPU-s. In particular, it took about 2 CPU hours on a single
processor to compute the diagrams in Eq. (\ref{1-2loop}) for a given set of
parameters $(ea,e\mu/m,m)$ for  $Q=0.8$. For any given set of $(Q,ea,m)$ we
chose 50 values for $0.5\le e\mu/m\le 2$ in such a manner that the points
between 0.98 and 1.35 were separated by 0.01.
For $m=1$ we took $a=0.0001,0.001,0.01,0.1,0.2,0.5,1.0$ and $Q=0.8,1.0,1.5$.
For $m=0.2,0.5,2.0$ and $5.0$ we took only the 4 larger values of $a$.

\newpage

\begin{figure}[ht]
\begin{center}
\psfig{file=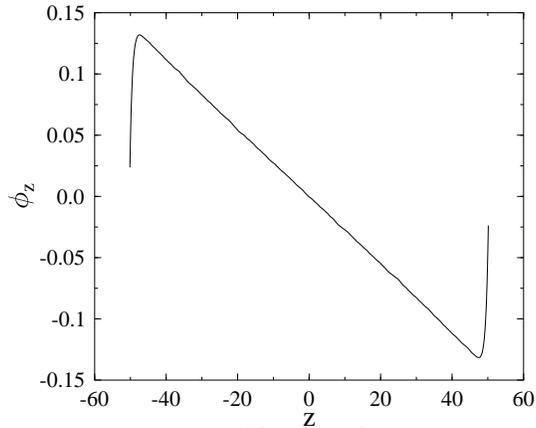,width=7cm}
\caption{The scalar field  $\phi_z$ for $b=0.025,~ m=e=1$.
The slope $b_s=-0.001078$ fitted in the central region should be compared to
$b_s=b e^2/(\kappa^2\pi)=-0.001074$ from Eq. (\ref{emsollin}).
\label{a0.02}}
\end{center}
\end{figure}

\begin{figure}[ht]
\begin{center}
\psfig{file=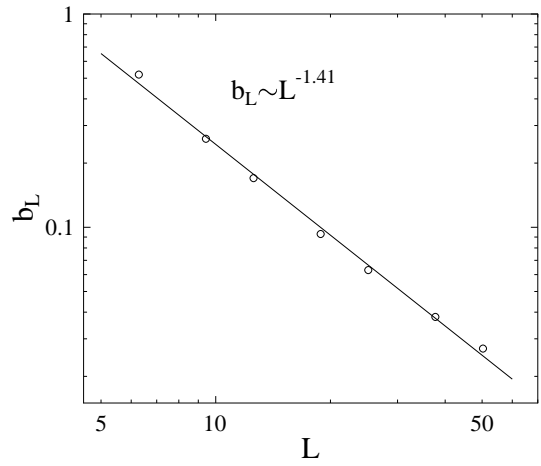,width=7cm}
\caption{Dependence of the point $b_{\mr{L}}$ on the size
$L$ of the system for $m=e=1$.
\label{linc}}
\end{center}
\end{figure}

\begin{figure}[ht]
\begin{center}
\psfig{file=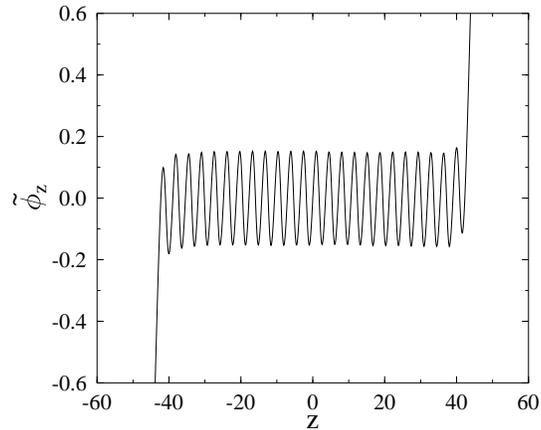,width=7cm}
\caption{The periodic part  $\tilde\phi_z$ of the scalar ground-state
field configuration for $b=0.5, m=e=1$.
\label{a0.5m1g1}}
\end{center}
\end{figure}

\begin{figure}[ht]
\begin{center}
\psfig{file=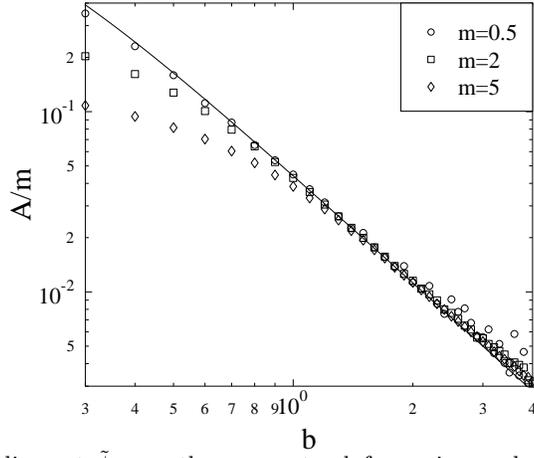,width=7cm}
\caption{Amplitude $A$ of the periodic part  
$\tilde\phi_z$ vs. the parameter $b$ for various values of the
electron mass $m$ and $e=1$. The solid line represents our perturbative
estimate for the amplitude given by Eq. (\ref{aext}).
The numerical values  justify the perturbative estimate for small values
of $A$.
\label{g1loga}}
\end{center}
\end{figure}

\begin{figure}[ht]
\begin{center}
\psfig{file=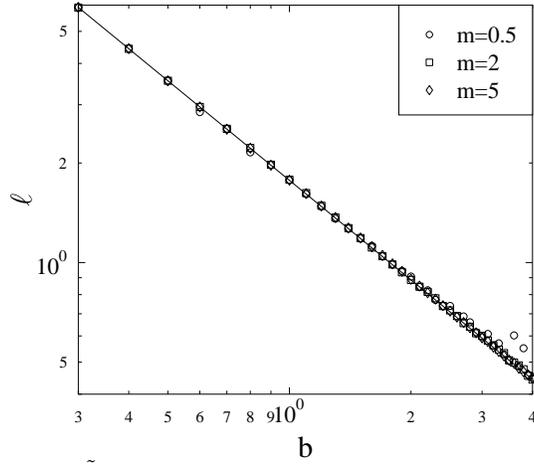,width=7cm}
\caption{Wavelength $\ell$ of the periodic part  
$\tilde\phi_z$ vs. the parameter $b$ for various values of the
electron mass $m$ and $e=1$. The solid line refers to the curve
$\ell=\sqrt{\pi}/b\equiv 1/\rho_\mr{ext}$.
\label{g1logl}}
\end{center}
\end{figure}

\begin{figure}[ht]
\begin{center}
\psfig{file=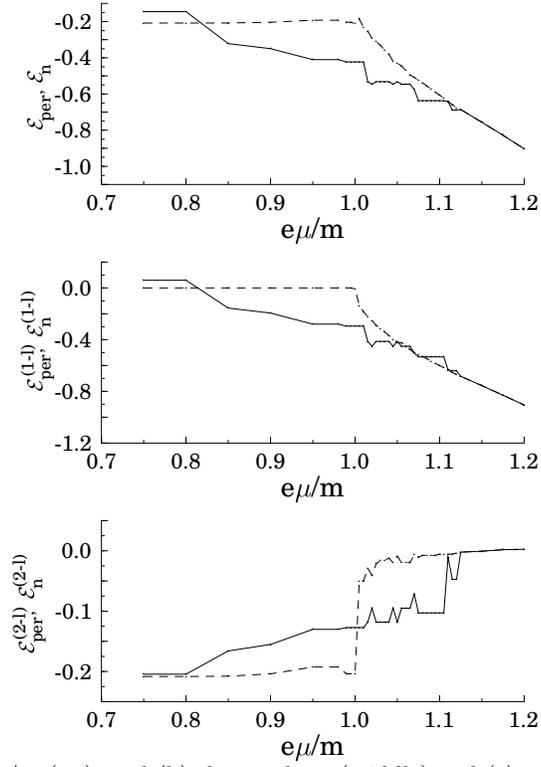,width=7cm}
\caption{(a) Energy densities vs. $e\mu/m$ (up), and (b) the one-loop (middle)
and (c) two-loop (down) contributions to those. The solid and dashed lines
refer to the periodic and the  homogeneous phases, respectively.
\label{1loop}}
\end{center}
\end{figure}

\begin{figure}[ht]
\begin{center}
\psfig{file=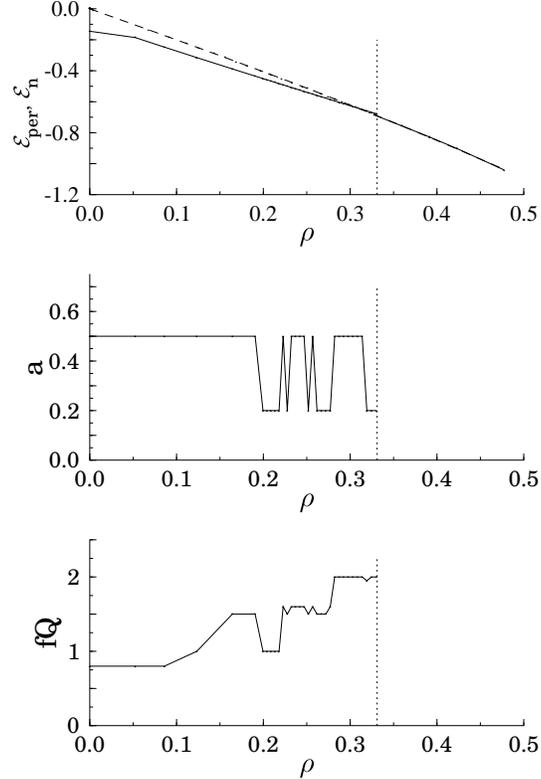,width=7cm}
\caption{Density dependences of (a) the energy density, (b) the amplitude $a$
of the photon condensate and (c) the product of the filling factor
$f$ and the wavenumber  $Q$ of the photon condensate
for $m=2$. The dotted lines indicate the `critical charge density' $\rho_c$,
solid and dashed lines correspond to the periodic and normal phases,
respectively.
\label{rhores}}
\end{center}
\end{figure}

\begin{figure}[ht]
\begin{center}
\psfig{file=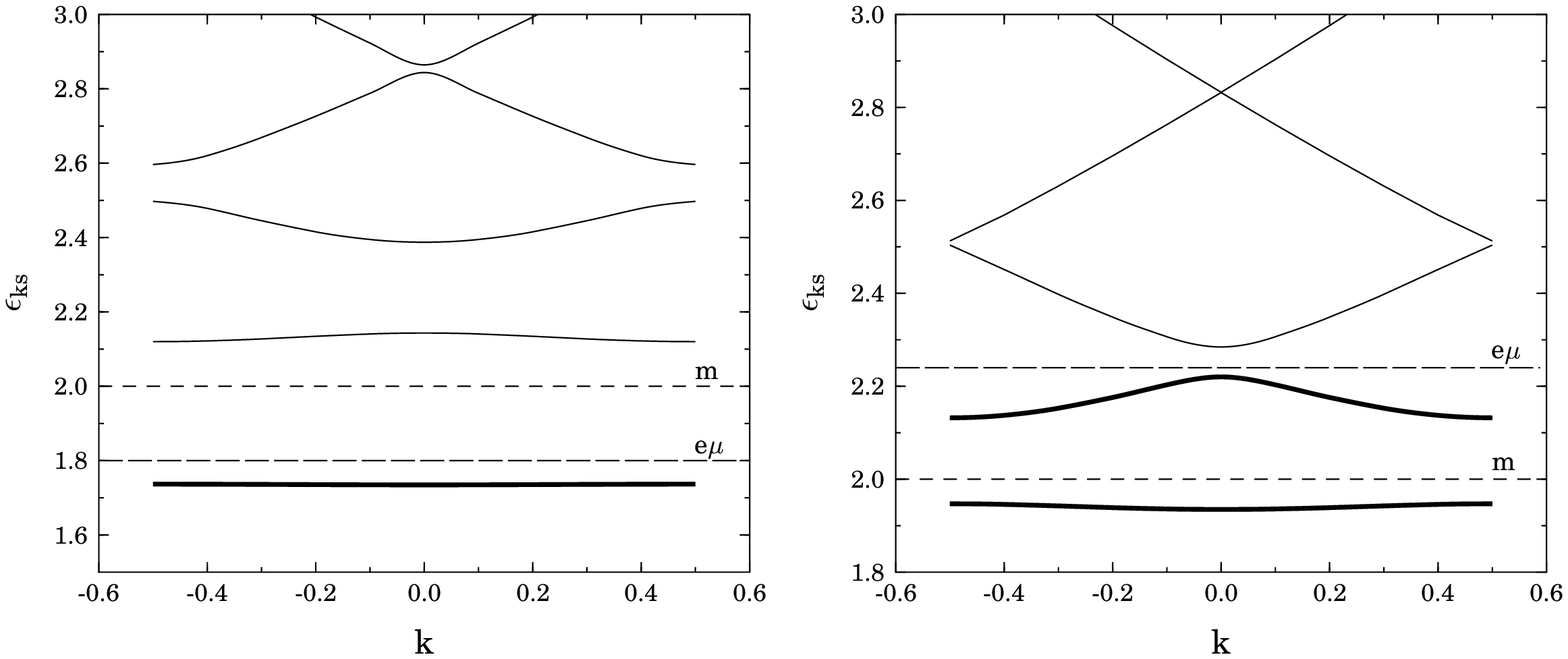,width=14cm}
\caption{Dispersion relations in the first Brillouin-zone with (a) one
occupied band (left) and (b) two occupied bands (right) sunk into the Dirac
sea for $ea=0.5$ and $ea=0.2$, respectively, $m=2$.
\label{brill}}
\end{center}
\end{figure}

\begin{figure}[ht]
\begin{center}
\psfig{file=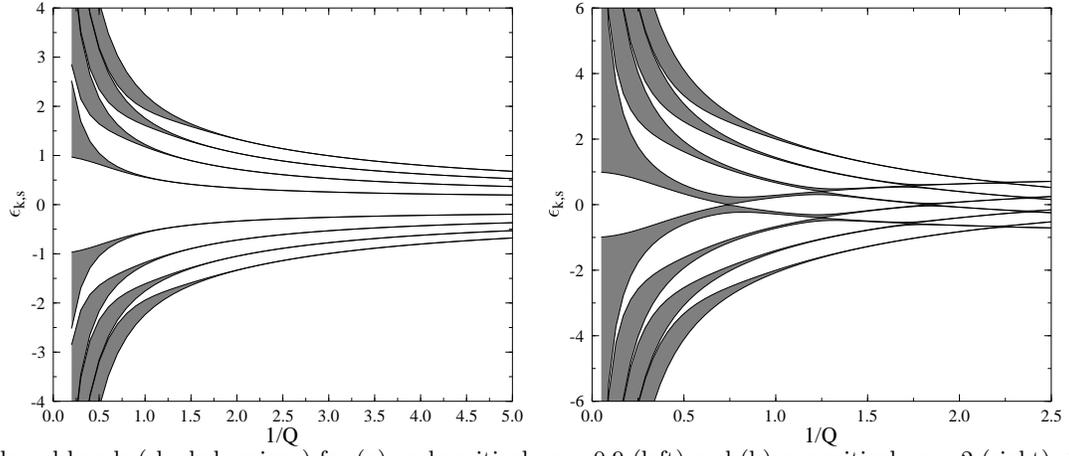,width=14cm}
\caption{Allowed bands (shaded regions) for (a) undercritical $ea=0.9$
(left)
and (b) overcritical $ea=2$ (right) static
periodic external fields with wavenumber $Q$ for $m=e=1$. 
\label{relslat}}
\end{center}
\end{figure}

\begin{figure}[ht]
\begin{center}
\psfig{file=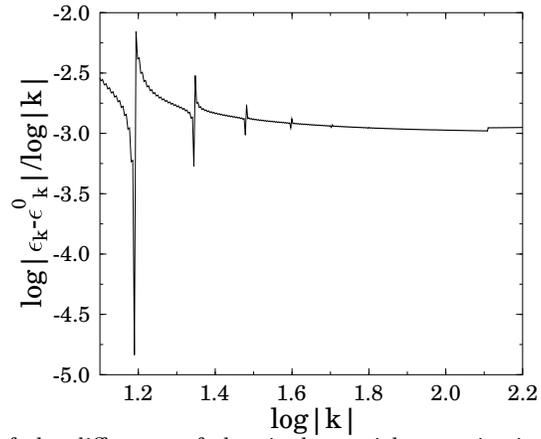,width=7cm}
\caption{Momentum-dependence of the difference of the single particle
energies in the expression (\ref{1l-sub}) of the one-loop energy
for $m=2$, $Q=1.1$ and $a=2$.
\label{konv}}
\end{center}
\end{figure}

\end{document}